\documentclass[12pt]{article}

\usepackage{times}
\usepackage{fullpage}
\usepackage{amsfonts,amssymb,amsmath,amsthm}
\usepackage{latexsym}
\usepackage{gauss}
\usepackage{tikz}
\usepackage{tikz-cd}
\usepackage{tkz-graph}
\usepackage{url}
\usepackage{thmtools}
\usepackage{thm-restate}
\usepackage{hyperref}
\usepackage[capitalise, nameinlink]{cleveref}
\usepackage{algorithm}
\usepackage{algorithmicx}
\usepackage{algpseudocode}
\usepackage{graphicx}
\usepackage{enumitem}
\newcommand{\N}{\mathbb{N}}

\newcommand{\R}{\mathbb{R}}

\newcommand{\zero}{\mathbf{0}}
\newcommand{\Id}{\mathbf{I}}
\newcommand{\vecu}{\vec{u}}
\newcommand{\Acal}{\mathcal{A}}

\newcommand{\Lcal}{\mathcal{L}}

\newcommand{\Mcal}{\mathcal{M}}
\newcommand{\M}{\mathcal{M}}
\newcommand{\Scal}{\mathcal{S}}
\newcommand{\Tcal}{\mathcal{T}}
\newcommand{\Fcal}{\mathcal{F}}
\newcommand{\Gcal}{\mathcal{G}}

\newcommand{\MINCUT}{\mathrm{MINCUT}}
\newcommand{\ARGMINCUT}{\mathrm{ARGMINCUT}}
\newcommand{\lin}{\mathrm{lin}}
\newcommand{\cut}{\mathrm{cut}}

\DeclareMathOperator*{\argmin}{argmin}

\newcommand{\laspan}{\mathrm{span}}
\newcommand{\cross}{\mathrm{overlap}}

\def\01{\{0,1\}}

\newcommand{\ith}{i^{\scriptsize \mbox{{\rm th}}}}

\newcommand{\cd}{\mathrm{cdim}}
\newcommand{\sep}{\mathrm{sep}}
\newcommand{\mer}{\mathrm{mer}}

\newcommand{\braket}[2]{\langle#1, #2\rangle}
\newcommand{\rk}{\mathrm{rk}}

\newcommand{\1}{\mathbf{1}}

\newtheorem{theorem}{Theorem}

\newtheorem{lemma}[theorem]{Lemma}
\newtheorem{corollary}[theorem]{Corollary}
\newtheorem{proposition}[theorem]{Proposition}
\newtheorem{remark}[theorem]{Remark}

\newtheorem{claim}[theorem]{Claim}
\newtheorem{fact}[theorem]{Fact}

\theoremstyle{definition}

\newtheorem{definition}[theorem]{Definition}

\title{On the cut dimension of a graph}
\date{}

\begin{document}
\author{Troy Lee\thanks{Centre for Quantum Software and Information, University of Technology Sydney. Email: troyjlee@gmail.com}
\and Tongyang Li\thanks{MIT,  Email: tongyang@mit.edu}
\and Miklos Santha\thanks{CNRS, IRIF, Universit\'e de Paris; Centre for Quantum Technologies and Majulab, National University of Singapore. Email: miklos.santha@gmail.com}
\and Shengyu Zhang\thanks{Tencent Quantum Laboratory. Email: shengyzhang@tencent.com}}
\maketitle

\begin{abstract}
Let $G = (V,w)$ be a weighted undirected graph with $m$ edges.  The cut dimension of $G$ is the dimension of the
span of the characteristic vectors of the minimum cuts of $G$, viewed as vectors in $\{0,1\}^m$. For every $n \ge 2$ we show that the
cut dimension of an $n$-vertex graph is at most $2n-3$, and construct graphs realizing this bound.

The cut dimension was recently defined by Graur et al.\  \cite{GPRW20}, who show that the maximum cut dimension of an $n$-vertex
graph is a lower bound on the number of cut queries needed by a deterministic algorithm to solve the minimum cut problem on $n$-vertex graphs.
For every $n\ge 2$, Graur et al.\  exhibit a graph on $n$ vertices with cut dimension at least $3n/2 -2$, giving the first lower bound larger than $n$
on the deterministic cut query complexity of computing mincut.  We observe that the cut dimension is even a lower bound on the number of \emph{linear}
queries needed by a deterministic algorithm to solve mincut, where a linear query can ask any vector $x \in \R^{\binom{n}{2}}$ and receives the answer $w^T x$.
Our results thus show a lower bound of $2n-3$ on the number of linear queries needed by a deterministic algorithm to solve minimum cut on $n$-vertex graphs,
and imply that one cannot show a lower bound larger than this via the cut dimension.

We further introduce a generalization of the cut dimension which we call the $\ell_1$-approximate cut dimension.  The $\ell_1$-approximate
cut dimension is also a lower bound on the number of linear queries needed by a deterministic algorithm to compute minimum cut.  It is always at least
as large as the cut dimension, and we construct an infinite family of graphs on $n=3k+1$ vertices with $\ell_1$-approximate cut dimension $2n-2$, showing
that it can be strictly larger than the cut dimension.
\end{abstract}

\newpage


\section{Introduction}
Let $G = (V,w)$ be a weighted undirected $n$-vertex graph where $w$ is an $\binom{n}{2}$-dimensional nonnegative real vector assigning a (possibly zero) weight to each edge slot.  For a
nontrivial subset $\emptyset \ne X \subsetneq V$, let $\Delta(X)$ be the set of edges of $G$ with one endpoint in $X$ and one endpoint in $\bar{X} = V \setminus X$.  A \emph{cut} $S$ in $G$ is a subset of edges of the form $\Delta(X)$ for
a nontrivial set $X$.  The sets $X$ and $\bar{X}$ are called the \emph{shores} of the cut.  For a cut $S$, its \emph{weight} is the sum of the weights of the edges in $S$, denoted $w(S)$.
The minimum cut problem is to find the minimum of $w(S)$ over all cuts $S$.  The study of algorithms for the minimum cut
problem in theoretical computer science goes back at least to the 1960's and has given rise to a vast and beautiful literature.
Minimum cut is also a problem of great practical importance with applications
to, for example, clustering algorithms and evaluating network reliability. Randomized algorithms can solve the minimum cut problem in nearly linear time: in 1996 Karger gave an
algorithm with running time $O(m \log^3(n))$ to compute the minimum cut of a weighted graph with $m$ edges \cite{Karger00}.  This was the best known bound until very recently when two independent
works improved on it.  Gawrychowski, Mozes, and Weimann \cite{GMW20} gave a randomized algorithm with running time $O(m\log^2(n))$ \cite{GMW20} and Mukhopadhyay and Nanongkai \cite{MN20} gave a randomized algorithm with time complexity $O(m\frac{\log^2(n)}{\log\log n}+n\log^6(n))$.   Gawrychowski, Mozes, and Weimann \cite{GMW20b} later improved the running time of the Mukhopadhyay and Nanongkai algorithm to $O(m\frac{\log^2(n)}{\log\log n}+n\log^{3+\varepsilon}(n))$.

For \emph{simple} graphs $G$, randomized algorithms are known with running times $O(m \log(n))$ and $O(m + n \log^3(n))$ \cite{GNT20}.  For simple graphs even nearly linear
time \emph{deterministic} algorithms are known.  Kawarabayashi and Thorup gave an $O(m \log^{12}(n))$ time algorithm \cite{KT19}, which was subsequently improved to
$O(m (\log(n) \log \log n)^2)$ by Henzinger, Rao, and Wang \cite{HRW20}.

Our work spans two aspects of the study of the minimum cut problem.  The first is to query complexity lower bounds on minimum cut.
A natural model in which to study the query
complexity of minimum cut is for algorithms allowed to make \emph{cut queries}.  A cut query algorithm can query any subset $\emptyset \ne X \subsetneq V$ and receives the
answer $w(\Delta(X))$.  One motivation to study cut query algorithms comes from submodular function minimization.  The cut function $f(X) = w(\Delta(X))$ is a submodular
function, and finding the minimum cut value is equivalent to finding the minimum value of $f$ over all nontrivial sets $X$.  The problem of minimizing a submodular function is often studied with respect to an evaluation
oracle, which in the case of the cut function is exactly a cut query.

Harvey \cite{Harvey08} observed that results on the deterministic communication complexity of deciding graph connectivity \cite{HMT88} imply that any deterministic cut query algorithm to compute minimum cut,
or even to decide if the graph is connected or not, must make at least $c n$ cut queries, for a constant $c < 1$.  Analogous results on the randomized communication complexity of connectivity \cite{BFS86}
imply an $\Omega(n/\log(n))$ lower bound on the number of cut queries needed by a randomized algorithm to compute minimum cut (or even connectivity).

On the algorithms side, Rubinstein, Shramm, and Weinberg \cite{RSW18} gave a randomized algorithm computing the minimum cut of a simple graph with $O(n \log(n)^3)$ many
cut queries\footnote{\cite{RSW18} state the bound as $\tilde O(n)$ but we estimate their sparsifier based algorithm to make $O(n \log(n)^3)$ many cut queries.}.
Recently, Mukhopadhyay and Nanongkai \cite{MN20} used a different approach based on Karger's 2-respecting tree algorithm
\cite{Karger00} to also give a randomized $\tilde O(n)$ cut query algorithm to compute minimum cut in a general undirected weighted graph.

For deterministic cut query algorithms, there remains a large gap between the best upper and lower bounds.  We are not aware of any deterministic algorithm for minimum cut
better than learning the entire graph, which can take $\Omega(n^2/\log(n))$ cut queries in the worst case.  On the lower bound side, Graur, Pollner, Ramaswamy, and Weinberg \cite{GPRW20} recently introduced
a very interesting lower bound technique called the \emph{cut dimension}, which we now describe.  Let $G = (V,w)$ be a weighted undirected graph with $n$ vertices and $m$ edges, and let $\M(G)$ be
the set of minimum cuts of $G$.  For a cut $S \in \M(G)$,
let $\chi(S) \in \{0,1\}^{m}$ be the characteristic vector of $S$
amongst the $m$ edges of $G$.  Let $\vec{\M}(G) = \{\chi(S): S \in \M(G) \}$.  The cut dimension of $G$, denoted
$\cd(G)$, is the dimension of $\laspan(\vec{\M}(G))$.
It is shown in \cite{GPRW20} that for any $n$-vertex graph $G$, the cut dimension $\cd(G)$ is a lower bound on the deterministic cut query complexity of computing
minimum cut on weighted $n$-vertex graphs.  Moreover, for every $n \ge 2$ they construct an $n$-vertex graph $G$ with cut dimension $3n/2-2$.

Besides showing lower bounds on cut query complexity, the cut dimension is a natural measure of the complexity of mincuts in a graph.  There is a rich
literature on the possible structure of mincuts in a graph.  Perhaps the first result of this kind is the cactus representation of mincuts by \cite{DKL76}.
A cactus for a graph $G$ is a sparse weighted graph $C$ that represents all the mincuts of $G$.  One consequence of the
cactus representation is that the number of possible mincuts in an $n$-vertex weighted graph is at most $\binom{n}{2}$.  This upper bound was later given
an algorithmic proof via Karger's famous contraction algorithm (\cite{Karger93}, Theorem~6.1).  The
$n$-vertex cycle graph has $\binom{n}{2}$ many minimum cuts and shows that this bound can be tight.

While the cycle has $\binom{n}{2}$ many mincuts, these cuts live in an $n$-dimensional space as the $n$-vertex cycle only has $n$ edges.  Is it possible to construct graphs
with many cuts that also have high cut dimension?  We show that this is not possible, and in fact the cut dimension of an $n$-vertex graph is at most $2n-3$.
\begin{theorem}[Main Upper Bound]
\label{thm:cd_upper}
For any weighted undirected graph $G$ on $n\ge 2$ vertices it holds that $\cd(G) \le 2n-3$.
\end{theorem}
Like the cactus representation, this shows another aspect in which the mincuts of a graph are constrained to have a relatively simple structure.  We further show that this bound is
tight by constructing graphs with cut dimension $2n-3$ for every $n \ge 2$.
\begin{theorem}[Main Lower Bound]
\label{thm:cd_lower}
For every $n \ge 2$ there exists an $n$-vertex weighted undirected graph $G$ with $\cd(G) = 2n-3$.
\end{theorem}

In addition to shedding further light on the structure of minimum cuts, this improves the best known lower bound on the deterministic cut query complexity of
the minimum cut problem to $2n-3$.  We additionally show that the cut dimension is even a lower bound on a stronger query model called the linear query model, recently
studied in \cite{ACK20}.  In the linear query model, the algorithm can query any vector $x \in \R^{\binom{n}{2}}$ and receives the answer $\braket{w}{x}$, the
inner product of $w$ and $x$.  Linear queries can be much more powerful than cut queries as one can completely learn an unweighted graph with a single linear query.
By an information theoretic argument learning an unweighted graph can require $\Omega(n^2/\log(n))$ many cut queries since each cut query reveals at most
$O(\log (n))$ bits.

We further introduce a lower bound technique which is a generalization of the cut dimension that we call the $\ell_1$-approximate cut dimension.  This technique
looks not just at mincuts in the graph, but all cuts.  We again look at the span of the dimension of these cuts with
an additional twist.  Suppose the weight of a minimum cut in $G$ is $\lambda$ and cut $S$ has $w(S) = \lambda + \delta$.  Abusing notation we will let $S$
represent both a set of edges and the characteristic vector $S \in \{0,1\}^{\binom{n}{2}}$ of $S$ among all edge slots.
The vector $S$ can be \emph{perturbed} to $S - u$ for
any vector $u \ge 0$ with $\|u\|_{1,w} \le \delta$.  Here $\| u \|_{1,w} = \sum_i |w(i) \cdot u(i)|$ is the $\ell_1$ norm of $u$ weighted by the edge weights of
the graph.  The $\ell_1$-approximate cut dimension of $G$ is then the minimum over all valid perturbations of the dimension of the span of the perturbed cut vectors.

The minimization over all perturbations makes the $\ell_1$-approximate cut dimension a difficult quantity to lower bound.  We are able to show, however, that the $\ell_1$-approximate cut
dimension can be strictly larger than the cut dimension.  For every $k \in \N$ and $n = 3k+1$, we construct an unweighted $n$-vertex graph $G$ whose $\ell_1$-approximate cut dimension is $2n-2$.
This has the following application.
\begin{theorem}
\label{thm:2n-2}
Any deterministic linear query algorithm that correctly computes the minimum cut of all $n$-vertex weighted undirected graphs must make
at least $2n-2$ queries in the worst case.
\end{theorem}

Computing the minimum cut of a graph with cut queries is a special case of finding the nontrivial minimum of a symmetric submodular function $f:2^V \rightarrow \R$ with evaluation queries.  That is,
to find $\min_{X: \emptyset \ne X \subsetneq V} f(S)$ for a submodular $f$ that satisfies $f(X) = f(V \setminus X)$ for all $X \subseteq V$.  As linear queries are more powerful than cut queries, \cref{thm:2n-2} also
implies a $2n-2$ evaluation query lower bound for a deterministic algorithm finding a nontrivial minimum of a symmetric submodular function, which is currently the best known.

\subsection{Techniques}
We give two different proofs of the $2n-3$ upper bound on the cut dimension and two different techniques to create graphs with cut dimension $2n-3$.
The first proof is direct and uses the \emph{combinatorial uncrossing} technique, and in particular a key lemma of Jain \cite{Jain01} in his factor of $2$ approximation
algorithm for the survivable network design problem.  The second proof is by induction and follows a framework for constructing a cactus representation of
the mincuts of a graph \cite{DKL76, FF09}.  The second proof uses very few properties of mincuts and seems better suited to also upper bound the $\ell_1$-approximate 
cut dimension, one of our main open questions.

Key to both proofs is the concept of when cuts \emph{cross} each other.  Two cuts $\Delta(X), \Delta(Y)$ are
said to \emph{cross} if all four of the intersections $X \cap Y, \bar{X} \cap Y, X \cap \bar {Y}, \bar{X} \cap \bar{Y}$ are non-empty.  Note that in the definition of crossing
it does not matter which shore we take to define the cut, thus crossing is a property of the cuts themselves.  A family $\Lcal$ of cuts is called \emph{cross-free}
if for all cuts $S, T \in \Lcal$ it holds that $S$ and $T$ do not cross.

In the first upper bound proof, we first show that any cross-free family of cuts has cardinality at most $2n-3$ (see \cref{sec:card}).  We then use Jain's lemma \cite{Jain01} (stated in \cref{thm:span}) to conclude
that for a maximal cross-free subset $\Lcal \subseteq \M(G)$ it holds that
$\vec{\Lcal} = \{\chi(S) : S \in \Lcal\}$ spans the set $\vec{\M}(G)$.
This shows that the cut dimension of a graph is at most $2n-3$.

In the first lower bound proof we use a tree-representation of a cross-free
family of cuts to show that in a complete graph the cut vectors of a cross-free family of cuts are linearly independent (\cref{prop:lin_ind}).  Thus the lower bound reduces
to constructing a graph whose minimum cuts are a cross-free family of cuts of size $2n-3$.  Such a construction has already been given by Chandra and Ram \cite{CR04}.  We go
a step further, however.  For any $\Lcal$ which is a cross-free family of cuts from a complete $n$-vertex graph with $|\Lcal| = 2n-3$,
in \cref{thm:explicit} we explicitly give the edge weights of a complete weighted graph $G$ such that $\M(G) = \Lcal$ and therefore $\cd(G) = 2n-3$.
This task is made easier by \cref{lem:min_value}, which states that if $\Lcal$ is a cross-free family of cuts of size $2n-3$ that all have the
same weight, then this must be the weight of a minimum cut of the graph.  This lemma is again shown by the combinatorial uncrossing technique.
This reduces the construction problem to solving the linear program of finding a positive vector $w$ that makes all cuts in $\Lcal$ have the same weight.  We
explicitly give a solution to this linear program by viewing it as a flow problem on the tree-representation of $\Lcal$.

The second upper bound proof is by induction and follows methods to construct a cactus representation of mincuts \cite{DKL76, FF09}.  In the base case $n=2$
it is easy to see that the cut dimension is at most $2n-3=1$.  For the inductive step, when $G$ is an $n > 2$ vertex graph,
there are 3 cases to consider.  We call a cut of the form $\Delta(\{v\})$ a \emph{star cut}, and we will refer to all other cuts as  \emph{non-star} cuts.  The first case is where all cuts in $\M(G)$ are star cuts.
As the graph has $n$ vertices there are at most $n$ star cuts and so in this case the cut dimension is at most $n \le 2n-3$.
The second case is where for every non-star cut $S \in \M(G)$ there is a cut $T \in \M(G)$ which crosses $S$.  In this case \cite{DKL76} show that the graph
must be a cycle and the cut dimension is again at most $n \le 2n-3$.

The interesting case is where there is a non-star cut $\Delta(V_0) \in \M(G)$ which is not crossed by any other cut in $\M(G)$.  Let $V_1 = \bar{V}_0$. In this case we use a decomposition of $G$ along the cut
$\Delta(V_0)$, that we call the {\em separation} of $G$, into two smaller graphs $G_b$, for $b \in \{0,1\}$.
The graph $G_b$ is formed from $G$ by contracting $V_{1-b}$ into a single new vertex $v_{1-b}$.
We show that
$\cd(G) \le \cd(G_0) + \cd(G_1) -1$ which implies immediately the upper bound.
Indeed, let $k = |V_0| \ge 2$.
Then $G_0$ is a graph on $k+1$ vertices and $G_1$ is a graph on $n-k+1$ vertices, both of which are less than $n$.  The inductive hypothesis
therefore gives $\cd(G) \le 2(k+1)-3 + 2(n-k+1)-3 -1 = 2n-3$.

For the second lower bound proof we use the {\em merge} operation
which creates from two graphs $G_b$, for $b \in \{0,1\}$, and a specified vertex $v_{1-b}$ from each,
a composed graph $G$ where the vertices $v_0, v_1$ are not present but the cut $\Delta(V_0)$ reflects
the structure of the star cuts at $v_0$ and $v_1$ in the original graphs.
The operations separation and
merge are inverses in the sense that if we apply merge to $\{G_0, G_1\}$ followed by separation on the
resulting graph $G$, we receive back $\{G_0, G_1\}$. We also show that
the inequality $\cd(G) \le \cd(G_0) + \cd(G_1) -1$ holds with equality if  $\Delta(V_0)$ is a connected graph. This enables us to construct inductively a sequence of graphs $G^{(n)}$ on $n$ vertices
whose cut dimension is $2n-3$. In the base case $G^{(3)}$ is the complete graph on 3 vertices where all
the edges have the same weight. Then $G^{(n)}$ is defined as the merge of $G^{(3)}$ and $G^{(n-1)}$
where the specified vertices can be chosen arbitrarily. Since the separation of $G^{(n)}$ along
the newly constructed complete cut gives back $G^{(3)}$ and $G^{(n-1)}$, from the inductive hypothesis
we conclude that $\cd(G) = \cd(G_0) + \cd(G_1) -1 = 2n-3.$

As the cut dimension is at most $2n-3$, we have to look to other methods in order to show larger lower bounds, if possible.  We propose
a generalization of the cut dimension which we call the $\ell_1$-approximate cut dimension.  In order to motivate this, we quickly explain
why the cut dimension is a lower bound on the linear query complexity of mincut.
The main idea behind the cut dimension lower bound on query complexity is to answer all queries of the algorithm according to an $n$-vertex graph $G = (V,w)$.
Supposing the algorithm makes $k$ queries, we package these into a $k$-by-$\binom{n}{2}$ matrix $A$ whose rows are the query vectors.
If there is a cut $S \in M(G)$ which is not in the rowspace of $A$, then by the Fredholm alternative there is a vector $z$ such that $Az= \zero$, where $\zero$ is the all-zero vector, but $\braket{S}{z} > 0$
and furthermore $z(i)=0$ whenever $w(i) = 0$.  Thus for a sufficiently small $\varepsilon > 0$ we have that $w - \varepsilon z \ge \zero$ and so
$G' = (V,w-\varepsilon z)$ defines a valid non-negatively weighted graph that has
all the same answers to the queries of the algorithm as $G$.  On the other hand, the weight of a minimum cut in $G'$ is strictly smaller than that of $G$ and
thus as the algorithm cannot distinguish $G$ and $G'$ it cannot correctly compute the weight of a minimum cut in all $n$-vertex graphs.

The $\ell_1$-approximate cut dimension extends this adversary argument to include \emph{all} the cuts of $G$ instead of just the mincuts.  If the
minimum cut weight of $G$ is $\lambda$ and $S$ is a cut with weight $\lambda+\delta$, then the algorithm will still fail if there is a $z$ such that
\begin{enumerate}
\item $w-z \ge \zero$
\item $Az = \zero$
\item $\braket{S}{z} > \delta$.
\end{enumerate}
The reason is the same: the graph $G' = (V,w-z)$ has all the same answers to the queries made by the algorithm as $G$ yet has
a cut with weight strictly smaller than $\lambda$.

Taking the dual of the corresponding linear program shows that such a vector $z$ will not exist iff $S-u$ is in the rowspace of $A$
for a vector $u \ge 0$ with $\|u\|_{1,w} \le \delta$.  This leads us to define the $(w,c)$ one-sided row-by-row $\ell_1$ approximate rank of a matrix.
For a matrix $Y \in \R^{M \times N}$
this is defined by a weight vector $w \in \R^N$ and a cost vector $c \in \R^M$ with $c \ge \zero$.  It is the minimum rank of a matrix
$\tilde Y$ such that $\tilde Y \le Y$ and $\|Y(i,:) - \tilde Y(i,:)\|_{1,w} \le c(i)$ for every row $i$, where $Y(i,:)$ denotes the $\ith$ row of $Y$.
Let $G = (V,w)$ be a graph and the weight of a minimum cut in $G$ be $\lambda$.
The $\ell_1$-approximate cut dimension of a graph $G = (V,w)$, denoted $\widetilde{\cd}(G)$,
is the $(w,c)$ one-sided row-by-row $\ell_1$-approximate rank of the matrix $Y$ whose rows are the vectors
$S \in \{0,1\}^{\binom{n}{2}}$ for every cut $S$ of $G$, and where $c = Yw - \lambda \1$, and $\1$ is the all-one vector.

Lower bounding the rank under such an $\ell_1$ perturbation is a difficult task.  However, we are able to show an infinite family
of graphs whose $\ell_1$-approximate cut dimension is $2n-2$, thereby showing the $\ell_1$-approximate cut dimension can be strictly larger than the cut dimension.
This lower bound is of a ``direct sum'' type.  We show that the $\ell_1$-approximate cut dimension of $K_4$, the complete
graph on $4$ vertices, is $6$, giving a tight lower bound of $6$ on the number of linear queries needed to compute minimum
cut on a 4 vertex graph.  We then show that the direct union (see \cref{def:du}) of
$k$ copies of $K_4$ has $\ell_1$-approximate cut dimension $6k$.  The proof is tailored to the specific properties of the cut vectors of $K_4$,
and makes use of Gaussian elimination and properties of diagonally dominant matrices.

Related to the $\ell_1$-approximate cut dimension is the question of the cut dimension of approximate mincuts.  For $\alpha \ge 1$ call a cut $S$
of a graph $G$ an $\alpha$-near-mincut if its weight is at most $\alpha$ times the weight of a minimum cut of $G$.  Let $\M_{\alpha}(G) = \{S : S \mbox{ is an } \alpha\mbox{-near-mincut of } G\}$.
It is known that $|\M_\alpha(G)| \le \binom{n}{2}$ for $\alpha < 4/3$ \cite{NNI97} (see also the beautiful proof given in Theorem~15 of \cite{GR95}).
Even for $\alpha < 3/2$ the number of $\alpha$-near-mincuts is $O(n^2)$ \cite{HW96}, which is a sharp threshold as there exist graphs with $\Omega(n^3)$ many $3/2$-mincuts.
There is also a generalization of the cactus representation of mincuts in terms of a tree of deformable polygons that applies to $\alpha$-near-mincuts for $\alpha < 6/5$ \cite{BG08}.
 in \cref{sec:alphacuts} we show that if $G$ is a \emph{simple} graph then $\dim(\laspan(\vec{\M}_\alpha(G))) = O(n)$ for any $\alpha < 2$ (\cref{thm:alpha}).  This bound is tight as for $\alpha = 2$ the
 unweighted complete graph $K_n$ witnesses $\dim(\laspan(\vec{\M}_2(K_n))) = \binom{n}{2}$.  For weighted graphs, on the other hand, we show that for any $\alpha > 1$ there exists an $n$-vertex
 weighted graph $G$ with $\dim(\laspan(\vec{\M}_\alpha(G))) = \binom{n}{2}$.

\subsection{Open Problems}
Several interesting open problems remain from this work.
\begin{itemize}
  \item There is still a large gap between the known upper and lower bounds on the deterministic cut/linear query complexity of minimum cut.
  What is the right answer?  We conjecture there is a deterministic cut query algorithm for minimum cut making $O(n^{2-\varepsilon})$ many queries
  for some $\varepsilon > 0$.
  \item Is the $\ell_1$-approximate cut dimension $O(n)$ for any $n$-vertex graph?  Also can one show a general direct sum theorem for the $\ell_1$-approximate
  cut dimension?
\end{itemize}

\subsection{Organization}
The rest of the paper is organized as follows. We review necessary backgrounds about graphs, operations on graphs, and query models in \cref{sec:prelim}. In \cref{sec:min-query}, we show that the cut dimension is a lower bound on the deterministic linear query complexity of computing minimum cut. We then prove that the cut dimension is at most $2n-3$ in \cref{sec:upper}, and give an explicit construction of graphs with cut dimension $2n-3$ in \cref{sec:construction}. In \cref{sec:another-proof}, we give another proof for both the upper and lower bounds on $2n-3$ using graph operations.  In \cref{sec:l1-approx} we show a $2n-2$ lower bound on $\ell_1$-approximate cut dimension which implies \cref{thm:2n-2}.  Finally, in
\cref{sec:alphacuts} we show that for a simple graph $G$ and $1 \le \alpha < 2$ it holds that $\dim(\laspan(\vec{\M}_\alpha(G))) = O(n)$.


\section{Preliminaries}\label{sec:prelim}

For every natural number $n$, we denote by $[n]$ the set
$\{1,2, \ldots, n\}$.
For a vector $z \in \R^n$ we write $z \geq \mathbf{0}$ if every coordinate of the vector
is at least 0, and similarly we write $z = \mathbf{0}$ if $z$ is the all-zero vector.
We denote the scalar product of two vectors $z,z' \in \R^n$
by $\braket{{z}}{z'}$. For any matrix, denote the rank of $A$ by $\rk(A)$.
We denote the disjoint union of sets $X$ and $Y$ by $X \sqcup Y.$

\subsection{Graphs, cuts, sets}

An undirected {\em weighted graph} on $n$ vertices is a couple
$G = (V,w)$, where $V$ is the set of vertices with $|V| = n$,
the set of edge slots
$V^{(2)}$ is the
set of subsets of $V$ with cardinality 2, and the weight function
$w : V^{(2)} \rightarrow \R $ is non-negative. We refer to the vertex set of $G$ as $V(G)$.
The set of edges of $G$
is defined as $E = \{e \in V^{(2)} : w(e) > 0\}$.
When in a graph $G=(V,w)$ the weight of every edge is 1, we say that the graph is {\em unweighted},
and we refer to it also as $G=(V,E)$; such graph is also called a {\em simple graph}.
For an edge $e = \{u,v\} $, we say that $u$ and $v$ are the endpoints of $e$.
For a subset $X \subseteq V$ of the vertices, we denote by $E(X)$ the set of edges in $E$
which have both endpoints in $X$, and for disjoint subsets $X, Y \subseteq V$, we denote by
$E(X, Y)$ the set of edges with exactly one endpoint in each of the two sets.
We extend the weight function $w$ to any subset $E'$ of the edges by $w(E') = \sum_{e \in E'} w(e)$.
We will deal only with graphs which have at least $2$ vertices.

We fix an ordering
$v_1 < v_2 < \cdots < v_n$ of the vertices which induces also an ordering $\{v_1, v_2\}, \{v_1, v_3\}, \ldots, \{v_{n-1},v_n\}$ of the edge slots as well as an ordering $e_1 < e_2 < \ldots < e_m$ of the $m=|E|$ edges.  We view $w \in \R^{\binom{n}{2}}$ as a vector whose $\ith$ coordinate gives the (possibly zero) weight of the $\ith$ edge slot
according to this ordering,
and we define $\vec{w} \in \{0,1\}^m$ as the restriction of $w$ to the edges.
With some slight abuse of notation, for a set of edges $S \subseteq E$, we use the same symbol $S$ to also denote
the characteristic vector in $\{0,1\}^{\binom{n}{2}}$ of $S$ among all edge slots.
We further need the characteristic vector of $S \subseteq E$ among the $m$ edges $E$, for which we use the
notation $\chi(S) \in \{0,1\}^m$.
For a family ${\cal F}$ of subsets of the edges, we use the notation
$\vec{{\cal F}} = \{ \chi(S) \in \{0,1\}^m : S \in {\cal F}\}$.

For $X \subseteq V$, we denote by $\bar{X}$ the set $V \setminus X$.
A {\em cut} $S$ is a set $E(X, \bar{X})$ for some $\emptyset \neq X \subsetneq V$.
We call
$X$ and $\bar{X}$ the {\em shores} of $S$, and we denote the cut by $\Delta(X)$.
A cut is a {\em star cut} if one of its shores is a singleton,
otherwise it is {\em non-star} cut. If the singleton shore of a star cut $S$ is $\{v\}$, then we say that $S$ is a star cut
at $v$.  The {\em weight} of a cut is the sum of the weights of its edges.
For a cut $S$ we define the {\em graph of the cut} $S$ as the unweighted graph $G(S)= (V', E')$
where $V'$ is the set of vertices in $V$
that are endpoints of at least one edge in $S$, and $E'=S$.
We say that a cut $S$ is
{\em connected} if $G(S)$ is a connected graph.
A cut is a {\em minimum} cut, or mincut, for short, if no other cut has smaller weight.
We denote by $\M(G)$ be the set of minimum cuts of $G$.
The {\em cut dimension} of $G$ is $\cd(G) = \dim(\laspan(\vec{\cal{M}}(G)))$.

Let $V$ be a set of size $n$. Two sets $X,Y \subseteq V$ are said to \emph{overlap} if $X \cap Y \ne \emptyset, \bar{X} \cap Y \ne \emptyset,
X \cap \bar{Y} \ne \emptyset$.
 A family $\Gcal$ of subsets of $V$ is said to be \emph{laminar} if for all $X,Y \in \Gcal$
it holds that $X$ and $Y$ do not overlap. A set family $\Gcal \subseteq 2^{V}$ is said to be \emph{closed under overlaps}
if for every $X,Y \in \Gcal$ that overlap it holds
that $X \cap Y, X \cup Y \in \Gcal$.  A laminar subset $\Lcal \subseteq \Gcal$ is said to be \emph{maximal in} $\Gcal$ if for every $X \in \Gcal - \Lcal$
there is a $Y \in \Lcal$ such that $X,Y$ overlap.  We say a laminar subset $\Lcal$ is maximal if it is maximal in $2^{V}$.

The sets $X,Y \subseteq V$ \emph{cross} if they overlap and additionally $\bar{X} \cap \bar{Y} \ne \emptyset$.
Note that if $X,Y$ cross then so do $X,\bar{Y}$. A set family $\Gcal \subseteq 2^{V}$ is said to be {\em cross-free} if for all $X,Y \in \Gcal$ it holds
that $X$ and $Y$ do not cross. Observe that if $X$ and $Y$ do not cross
then either $Y$ or $\bar{Y}$ is a subset of $X$ or $\bar{X}$.
Let $G = (V,w)$ be a graph with $n$ vertices.
Two cuts $\Delta(X)$ and $\Delta(Y)$ of $G$ are {\em crossing} if $X$ and $Y$ are crossing.
Let $\Fcal = \{ \Delta(X_1), \ldots,  \Delta(X_k) \}$ be a set of cuts of $G$.
We say that $\Fcal$ is {\em cross-free family of cuts} if
$\Gcal = \{X_1, \ldots, X_k\}$ is cross-free.  Note that it does not matter which shore we take to be in $\Gcal$.

There is a close relationship between cross-free families of cuts and laminar sets.
Let $\Fcal = \{ \Delta(X_1), \ldots,  \Delta(X_k) \}$ be a cross-free family of cuts where each $X_i \subseteq V$,
and let $X_i' = X_i$ if $v_1 \not \in X_i$ and $X_i' = \bar{X}_i$ otherwise.
The \emph{beach} of $\Fcal$ is the set $\Gcal = \{X_1', \ldots, X_k'\}$.
For a family of sets $\Gcal \subseteq 2^{V}$ we say that it is \emph{proper} if $\emptyset, V \not \in \Gcal$,
and we say that it is \emph{complement free} if it does not contain $X,Y$ with $Y = \bar{X}$.

\begin{claim}\label{claim:cross-free}
Let $\Fcal$ be a cross-free family of distinct cuts and $\Gcal$ its beach.  Then $\Gcal$ is proper, complement free and laminar.
\end{claim}

\begin{proof}
First, $\Gcal$ does not contain $\emptyset$ or $V$ because these are not shores of cuts.  It is complement free because $\Fcal$ contains
distinct cuts, and its beach contains exactly one representative shore from each cut.  Finally, we show that it is laminar.  Let
$X_1, X_2 \in \Gcal$.  By definition of a beach, neither of these sets contain $v_1$, thus $\bar{X}_1 \cap \bar{X}_2 \ne \emptyset$.  Therefore if
$X_1, X_2$ overlapped they would also cross, in contradiction to $\Fcal$ being a cross-free family of cuts.
\end{proof}

A mincut is {\em crossless} if no other
mincut crosses it. Observe that a star mincut is always crossless.
Also, if a  mincut $\Delta(X)$ is crossless then
for every mincut $\Delta(Y)$, either $Y$ or $\bar{Y}$ is a subset of $X$ or $\bar{X}$.
Crossing mincuts have a nice structural property which was already observed by~\cite{DKL76}.

\begin{claim}
\label{claim:mincuts}
Let $G=(V,w)$ be a weighted graph.
If $\Delta(X), \Delta(Y) \in \M(G)$ cross then
$\Delta(X \cap Y), \Delta(X \cup Y)  \in \M(G)$.
\end{claim}

\begin{proof}
We have $\Delta(X \cap Y) \neq \emptyset $ and $ \Delta(X \cup Y) \neq V$ because $\Delta(X)$ and $\Delta(Y)$ cross. The cut function is submodular therefore we have
\begin{equation*}
\label{eq:mincuts}
w(\Delta(X \cap Y) ) + w(\Delta(X \cup Y) ) \leq w(\Delta(X)) + w(\Delta(Y)).
\end{equation*}
Let $c$ be the weight of a minimum cut in $G$. Then the right hand side of the above inequality
is equal to $2c$, while its left hand side is at least $2c$. Therefore
$w(\Delta(X \cap Y) ) + w(\Delta(X \cup Y) ) = 2c$ from which the statement follows.
\end{proof}

\subsection{Operations on graphs}
We will use several operations on graphs.  The first of these is the \emph{direct union}.
\begin{definition}[direct union]
\label{def:du}
For two graphs $G_0= (V_0, w_0), G_1 = (V_1,w_1)$
with disjoint vertex sets,
and for vertices $v_0 \in V_0$ and $v_1 \in V_1$, the {\em direct union} of $G_0$ and $G_1$ at vertices
$v_0,v_1$ is the fusion of the two by identifying $v_0$ and $v_1$. Formally, the direct union is
$G_0^{v_0} \oplus   G_1^{v_1} = (V,w)$
where $V = ( V_0 \cup V_1 \cup \{v\} ) \setminus \{v_0,v_1\}$, for a new vertex
$v \not\in V_0 \cup V_1 $. The weight function  of $G_0^{v_0} \oplus   G_1^{v_1}$
is defined by
\begin{equation*}
w (\{x,y\}) =
\begin{cases}
w_b (\{x,y\}  & \text{ if } x,y \in V_b \setminus \{v_b\}, b\in \{0,1\}, \\
w_b(\{x,v_b\}) & \text{ if } x\in V_b \setminus \{v_b\}, y=v, b\in \{0,1\}, \\
0 &  \text{ otherwise }.
\end{cases}
\end{equation*}
\end{definition}
The cut dimension of a direct union is a simple function of the cut dimensions of its components.

\begin{claim}
\label{claim:directsum}
Let $G = G_0^{v_0} \oplus  G_1^{v_1}$ be the direct union of $G_0$ and $G_1$ at vertices
$v_0, v_1$.
Let $c_b$ be the weight of a minimum cut in $G_b$, for $b=0,1$.
Then $\cd(G) = \cd(G_0) + \cd(G_1)$ if
$c_0 = c_1$, and $\cd(G) = \cd(G_b)$ if $c_b < c_{1-b}$.
\end{claim}

\begin{proof}
Let $\Delta(X)$ be an arbitrary cut of $G$ where $v \not \in X$.
If $X \not \subseteq   V_b$,
for $b \in \{0,1\}$, then
the weight of the cut $\Delta(X)$ is at least $c_0 + c_1$, and therefore
it is not a minimum cut.  If $X \subseteq   V_b$, for some $b \in \{0,1\}$ then
the weight of $\Delta(X)$ in $G$ is the same as the weight of $\Delta(X)$ in $G_b$.
Therefore if $c_0 = c_1$ then every mincut in $G_0$ and
every mincut of $G_1$ is a mincut of $G$, and these are the only mincuts. Since their supports are disjoint,
we have $\cd(G) = \cd(G_0) + \cd(G_1)$. If $c_b < c_{1-b}$ then only the mincuts of $G_b$ are mincuts of $G$,
and therefore $\cd(G) = \cd(G_b)$.
\end{proof}

The next two operations, which are inverses of each other, give a decomposition of a graph along a cut into two smaller graphs, and a composition of two graphs into a bigger one by unfolding a star cut in
each components. The decomposition operation was essentially defined in~\cite{FF09}.
Let $G= (V,w)$ be a weighted graph and let $Z$ be a cut in $G$ with shores $X_0$ and $X_1 = V \setminus X_0$.
The {\em separation} of $G$ along the cut $Z$, denoted by $\sep(G,Z)$, is the set of two
graphs $\{ G_0= (V_0, w_0), G_1 = (V_1,w_1) \}$, where $V_b = X_b \cup \{v_{1-b}\}$, for
$b=0,1$ with new vertices $v_0, v_1$. The respective weight functions are defined by
$w_b (\{x,y\}) = w (\{x,y\}) $ for any $x,y \in X_b$, and $w_b (\{x,v_{1-b}\}) = \sum_{y \in V_{1-b}} w(\{x,y\})$ for any $x\in X_b$.

Let $G_0= (V_0, w_0), G_1 = (V_1,w_1)$ be two graphs on disjoint vertex sets, and let $v_b \in V_{1-b}$
be arbitrary vertices for $b \in \{0,1\}$. The {\em merge} of $G_0$ and $G_1$ along the vertices $v_1, v_0$,
denoted by $\mer( \{ (G_0, v_1), (G_1, v_0) \} )$,
is the graph $G=(V,w)$, where $V = (V_0 \cup V_1) \setminus \{v_0, v_1\}$. The weight function in $G$
is defined by $w (\{x,y\}) = w_b (\{x,y\})$ if $x,y \in V_b$, for $b \in \{0,1\}$, and
\[w (\{x,y\})  = w_0 (\{x,v_1\})  w_1 (\{v_0,y\}), \text{ if } x \in V_0 \text{ and }y \in V_1.\]
It follows from the definitions $\sep$ is the left inverse of $\mer$ if the star cut
at $v_1$ in $V_0$ and the star cut at $v_0$ in $V_1$ both have weight one,
and $\sep$ is the right inverse of $\mer$ if the weight of the cut $Z$ is one. We formally state
the former property.

\begin{claim}
\label{claim:sepmer}
Let $G_0= (V_0, w_0)$ and $ G_1 = (V_1,w_1)$ have disjoint vertex sets, and let $v_b \in V_{1-b}$ such that
$w_b(\Delta(v_{1-b})) =1$,
for $b=0,1$. Let $Z$ be the cut in $\mer( \{ (G_0, v_1), (G_1, v_0) \} )$ whose shores are
$V_0  \setminus \{v_1\}$ and $V_1 \setminus \{v_0\}$. Then $w(Z)=1$ and
$$
\sep ( \mer( \{ (G_0, v_1), (G_1, v_0) \} ), Z ) = \{ G_0, G_1 \}.
$$
\end{claim}

\subsection{Query models}
\begin{definition}[$\MINCUT_n$]
The input in the $\MINCUT_n$ problem is an $n$-vertex weighted undirected graph $G = (V,w)$.  The required output on $G$ is the weight
of a minimum cut in $G$.
\end{definition}
A deterministic algorithm correctly solves the $\MINCUT_n$ problem if it outputs the correct mincut weight for every $n$-vertex input graph $G$.
We consider algorithms given two models of query access to the input graph $G = (V,w)$, linear queries and cut queries.
A {\em linear query} for $G$ is a vector $x \in \R^{\binom{n}{2}}$,
and the query is answered by $\braket{x}{w}$.  A cut query is a vector $x \in \{0,1\}^{\binom{n}{2}}$
which is the characteristic vector of a cut in the complete $n$-vertex graph.  The answer
to a cut query is again $\braket{x}{w}$.  Clearly any cut query algorithm can be simulated by a linear
query algorithm.

We use $D_\cut(\MINCUT_n)$ to denote the minimum, over all deterministic query algorithms $\Acal$ that correctly solve $\MINCUT_n$,
of the maximum over all $n$-vertex input graphs $G = (V,w)$ of the number of cut queries made by $\Acal$ on $G$.
$D_\lin(\MINCUT_n)$ is defined analogously for linear queries.

Some authors instead define the output of the minimum cut problem to be a cut $S$ that achieves the minimum weight, rather than the weight itself.  Over $n$-vertex weighted
graphs let us denote this problem as $\ARGMINCUT_n$.  For linear and cut queries, an algorithm that finds a minimum cut $S$ can also return the weight of $S$ with one additional query.
Thus $D_{\lin, \cut}(\ARGMINCUT_n) \ge D_{\lin, \cut}(\MINCUT_n) - 1$, and the lower bounds we prove for $\MINCUT_n$ can be applied, minus 1, to $\ARGMINCUT_n$ as well.


\section{Lower bounds on the linear query complexity of MINCUT}\label{sec:min-query}
Graur et al.~\cite{GPRW20} introduce the cut dimension as a means to show lower bounds on the
deterministic cut query complexity of computing minimum cut.

\begin{theorem}[\cite{GPRW20}]
\label{thm:graur_lower}
If there is an $n$-vertex {weighted graph $G = (V,w)$} with $\cd(G) = k$ then $D_\cut(\MINCUT_n) \ge k$.
\end{theorem}

We show that this theorem even holds with respect to a stronger computational model where the algorithm is able to
make linear queries.  We also give a generalization of the cut dimension to a quantity which is at least as large,
and can be strictly larger, that we call the $\ell_1$-approximate cut dimension.  We now give an overview of the Graur et al.\  \cite{GPRW20}
argument in the context of linear queries and how we can extend it.

The proof of \cref{thm:graur_lower} is based on an adversary argument.  Suppose a deterministic cut query algorithm makes $k$ linear queries
and consider the execution of the algorithm on a fixed $n$-vertex graph $G = (V,w)$ whose set of minimum cuts is $\M(G)$.
Make a $k$-by-$\binom{n}{2}$ matrix ${A}$ whose rows are the query vectors asked by the algorithm. Suppose we can find a vector
${z} \in \R^{\binom{n}{2}}$ such that
\begin{enumerate}
  \item $w-{z} \ge \mathbf{0}$,
  \item ${A}{z} = \mathbf{0}$,
  \item There is a cut $S \in \M(G)$ such that $\braket{S}{z} > 0$.
\end{enumerate}
The existence of such a vector $z$ means the algorithm cannot correctly compute minimum cut
weight on all weighted $n$-vertex graphs.  The reason is that $G' = (V,w-z)$
is a valid non-negatively weighted graph by (1), has the same answers on all queries asked by the algorithm by (2),
and by (3) has minimum cut weight at most $\braket{S}{w-z} = \braket{S}{w} - \braket{S}{z} < \braket{S}{w}$,
which is strictly less than the minimum cut weight of $G$. As with $k$ queries the algorithm cannot distinguish whether the input is $G$ or $G'$, it cannot correctly output the minimum cut weight for all $n$-vertex weighted graphs.

A weaker condition than~(3) suffices for this argument to work.  Suppose that the minimum cut weight in $G$ is $c^*$.
Then the argument still goes through with the condition
\begin{itemize}
  \item[3'.] There is a cut $S$ such that $\braket{S}{z} > \braket{S}{w}-c^*$.
\end{itemize}
This is because the algorithm cannot distinguish the graph $G$ with minimum cut weight $c^*$ from the graph $G' = (V,w-z)$
which has minimum cut weight at most $\braket{S}{w-z} < c^*$.

In order to understand what kind of bound this argument gives, for fixed $w,A,S$ we define the quantity $\alpha(w,A,S)$
which is given by the following linear program.
\begin{equation*}
\begin{aligned}
\alpha(w,A,S) =\  & \underset{z}{\text{maximize}}
& & \braket{S}{z} \\
& \text{subject to}
& & w-z \ge 0 \\
& & & Az = \zero
\end{aligned}
\end{equation*}
Taking the dual of this program gives
\begin{equation*}
\begin{aligned}
\alpha(w,A,S) =\  & \underset{v}{\text{minimize}}
& & \braket{S - A^Tv}{w} \\
& \text{subject to}
& & S - A^T v \ge 0
\end{aligned}
\end{equation*}

The dual tells us that a vector $z$ having large overlap with $S$ and satisfying items $(1),(2)$ above exists
iff the vector $S$ is \emph{far away} from the rowspace of $A$.  The notion of far away here is a one-sided $\ell_1$ distance
weighted by $w$.  It is one-sided because the condition $S - A^T v \ge 0$ tells us we are looking to approximate $S$ by
vectors in the rowspace of $A$ that are entrywise at most $S$.  As $S - A^T v \ge 0$ and $w \ge 0$ this means
$\braket{S - A^Tv}{w} = \sum_i | w(i) \cdot (S (i)- A^Tv)| = \|S - A^Tv \|_{1,w}$, where
$\|u\|_{1,w}$ is defined to be $\sum_i |u(i)w(i)|$.  Thus the value of the dual can be interpreted as the one-sided $\| \cdot \|_{1,w}$ distance
between $S$ and the rowspace of $A$.

This leads us to define an $\ell_1$ approximate version of the cut dimension.  The notion
we need is given by the following definitions.
\begin{definition}[one-sided row-by-row $\ell_1$-approximate rank]
Let $Y \in \R^{M \times N}$ be a matrix, $w \in \R^N$ a weight vector and
$c \in \R^M$ a cost vector.  We define the $(w,c)$ one-sided row-by-row $\ell_1$-approximate rank of $Y$ to be
the minimum rank of a matrix $\tilde Y$ such that $\tilde Y \le Y$ and $\| Y(i,:) - \tilde Y(i,:) \|_{1,w} \le c(i)$,
for all $1 \leq i \leq M$.
\end{definition}

\begin{definition}[$\ell_1$-approximate cut dimension]
Let {$G = (V,w)$} be an $n$-vertex weighted undirected graph with minimum cut weight $c^*$.
Let $M$ be $(2^{n-1}-1)$-by-$\binom{n}{2}$ matrix whose rows are $S \in  \{0,1\}^{\binom{n}{2}}$ for all cuts $S$ of $G$.
Let $c = Mw - c^*\1$, where $\1$ is the all one vector.  Then the $\ell_1$-approximate cut dimension of $G$, denoted $\widetilde{\cd}(G)$, is
the $(w,c)$ one-sided row-by-row $\ell_1$-approximate rank of $M$.
\end{definition}

\begin{theorem}
\label{thm:approx_cut}
If there is an $n$-vertex graph {weighted graph $G = (V,w)$} with $\widetilde{\cd}(G) = k$
then $D_\lin(\MINCUT_n) \ge k$.
\end{theorem}

\begin{proof}
Let $G = (V,w)$ be a graph with $\widetilde{\cd}(G) = k$ and let $c^*$ be the minimum cut weight of $G$.  Suppose for
contradiction there is a deterministic $k-1$ linear query algorithm that correctly computes the minimum cut of any
$n$-vertex graph.  Run this algorithm answering queries according to $G$ and
package the queries into a $(k-1)$-by-$\binom{n}{2}$ matrix $A$.

As the algorithm is correct, for every cut $S$ of $G$ it must be the case that $\alpha(w,A,S) \le \braket{S}{w} -c^*$.  If not, the graph $G' = (V,w-z)$, where $z$ is an optimal
solution to the primal of $\alpha(w,A,S)$, has minimum cut weight strictly smaller than $c^*$, yet $G'$ cannot be distinguished from $G$ by the algorithm.
Thus by the dual formulation of $\alpha(w,A,S)$, this means that for every cut $S$ of $G$ there is a vector $\tilde{S}= A^T v$ in the rowspace of $A$ such
that $\tilde S \le S$ and $\|S-\tilde{S}\|_{1,w} \le \braket{S}{w} -c^*$.  The matrix $\tilde M$ whose rows are $\tilde S$ for all
cuts $S$ therefore witnesses that $\widetilde{\cd}(G) \le \rk(A) \le k-1$, a contradiction.
\end{proof}

\begin{lemma}
\label{lem:cdless}
For any weighted graph $G = (V,w)$ we have $\cd(G) \le \widetilde{\cd}(G)$.
\end{lemma}

\begin{proof}
Suppose that $G = (V,w)$ has minimum cut weight $c^*$, and let $\M(G)$ be the set of minimum cuts of $G$.
Let $M$ be the $(2^{n-1}-1)$-by-$\binom{n}{2}$ matrix whose rows are $S \in \{0,1\}^{\binom{n}{2}}$ for all cuts $S$ of $G$
and let $c = Mw-c^*$.

Let $Y$ be the submatrix of $M$ where rows are restricted to cuts in $\M(G)$ and columns are restricted
to the edge slots $e$ where $w(e) > 0$.  Thus the rows of $Y$ are exactly the vectors $\chi(S)$ for $S \in \M(G)$.
and the rank of $Y$ is $\cd(G)$.  Any matrix $\tilde M$ which satisfies $\tilde M \le M$ and $\|M(i,:)-\tilde{M}(i,:)\|_{1,w} \le c(i)$
for all $i$ must contain $Y$ as a submatrix, as $c(i)=0$ for rows $i$ that correspond to minimum cuts and $w$ is positive on the
edge slots labeling the columns of $Y$.  Thus $\rk(\tilde M) \ge \rk(Y)$ for any $(w,c)$ one-sided row-by-row $\ell_1$ approximation
$\tilde M$ of $M$, giving the lemma.
\end{proof}

In \cref{sec:approx_construction} we will see that $\widetilde{\cd}(G)$ can be strictly larger than $\cd(G)$.  From \cref{thm:approx_cut} and
\cref{lem:cdless} we obtain the following corollary.
\begin{corollary}
\label{cor:cd_linear}
If there is an $n$-vertex weighted graph $G = (V,w)$ with $\cd(G) = k$
then $D_\lin(\MINCUT_n) \ge k$.
\end{corollary}


\section{The cut dimension is at most $2n-3$}\label{sec:upper}
In this section we prove \cref{thm:cd_upper} that $\cd(G) \le 2n-3$ for any undirected weighted graph $G$ on $n \ge 2$ vertices.  This will follow from
two facts:
\begin{enumerate}
  \item For $n \ge 2$ a cross-free family of cuts in an $n$-vertex graph has cardinality at most $2n-3$.
  \item If $\Lcal \subseteq \M(G)$ is a maximal cross-free subset of the mincuts of $G$
  then $\laspan(\vec{\Lcal}) = \laspan(\vec{\M}(G))$.
\end{enumerate}
We remind the reader that $\vec{\Lcal} = \{\chi(S) : S \in \Lcal\}$ where $\chi(S) \in \{0,1\}^{|E|}$ is the characteristic
vector of the cut $S$ amongst the edges of $G$.

These two facts are presented in the next two subsections.

\subsection{Cardinality of a cross-free family of cuts}
\label{sec:card}
Recall from \cref{claim:cross-free} that if $\Lcal$ is a cross-free family of cuts then the beach $\Gcal$ of $\Lcal$ is a laminar family of sets.
A standard inductive proof shows that a laminar family of subsets of a universe of cardinality $n$ that contains no singletons has size at most $n-1$, and thus a laminar
family in general has size at most $2n-1$.  A beach has the additional properties of being proper and complement free
which allows one to prove an upper bound of $2n-3$.  This is mentioned by \cite{Goemans06} in the paragraph after Theorem 4
under the heading ``Size of a Laminar Family'', who observes that the standard inductive proof also
implies the bound is attained only if the family includes the universe and at least one set and its complement.
See also Corollary~2.15 of \cite{KV18}, where it is shown that a proper laminar family has cardinality at most $2n-2$.

\begin{lemma}
\label{lem:laminar_size}
Let $n \ge 2$, $V$ a set of cardinality $n$, and $\Gcal \subseteq 2^{V}$ be a family of sets which is proper and laminar.  Then
$|\Gcal| \le 2n-2$.  If $\Gcal$ is proper, laminar, and complement free then $|\Gcal| \le 2n-3$.
\end{lemma}

\begin{proof}
First we show the $2n-2$ upper bound.
We prove by induction.  Consider first the base case where $n=2$ and $V = \{v_1, v_2\}$.  As $\emptyset, V \not \in \Gcal$ the only possible elements to include in $\Gcal$
are $\{v_1\}, \{v_2\}$ and $|\Gcal| \le 2 = 2n-2$.

Now we assume the statement is true for families of sets on a universe of $n-1$ elements and show it holds for families of sets on a universe of size $n$.  Let
$\Gcal \subseteq 2^{V}$ be a proper laminar family.  We say that $X \in \Gcal$ is maximal
if there is no set $Y \in \Gcal$ with $X \subset Y$.  Let $X_1, \ldots, X_m$ be the maximal sets in $\Gcal$.  Note that we must have
$X_i \cap X_j = \emptyset$ for all $i\ne j \in [m]$.  This is because for distinct maximal sets $X_i - X_j, X_j - X_i \ne \emptyset$ thus if
$X_i \cap X_j \ne \emptyset$ they would be overlapping.  If $\cup_{i=1}^m X_i \subsetneq V$ then the result already holds by the induction hypothesis.
Thus we may assume $m \ge 2$ and $X_1, \ldots, X_m$ form a partition of $V$.
The family $\Fcal_1 = \{Y : Y \subsetneq X_1\}$ is a laminar family on the universe $X_1$ which does not contain $X_1$.  Hence by the induction hypothesis it has at most $2|X_1|-2$ many sets.  This holds for all
$i =1, \ldots, m$, thus including $X_1, \ldots, X_m$ the total number of sets is $\sum_{i=1}^m 2|X_i| - m \le 2n-2$.

Now we show the $2n-3$ upper bound additionally assuming the family is complement free.  We show this result directly using the upper bound
of $2n-2$ we have just shown on the size of proper laminar families.  Let $\Gcal \subseteq 2^{V}$ be proper, laminar, and complement free,
and let $X_1, \ldots, X_m$ be the maximal sets in $\Gcal$, which again must be disjoint.  The number of subsets strictly contained in $X_i$ is at most $2|X_i| - 2$ by the previous result.  Thus, including $X_1, \ldots, X_m$
we can upper bound the size of $\Gcal$ by $\sum_{i=1}^m 2|X_i| - m$.  If $m > 2$ then the upper bound of $2n-3$ already holds.  If $m=1$ then as $\Gcal$ is
a proper family we must have $|X_1| \le n-1$ in which case the upper bound of $2n-3$ holds as well.  Finally, consider the case $m=2$.  In this case, if $|X_1 \cup X_2| < n$ then
the bound already holds.  If $X_1 \cup X_2 = V$ then $X_2 = \bar{X}_1$ and we must exclude one of these sets, giving a bound of $2n-2-1 = 2n-3$.
\end{proof}

\begin{remark}
From the proof in the proper, laminar, complement-free case we can observe for what maximal sets equality in the upper bound can hold.  The first is the case where there are three maximal sets
$X_1, X_2, X_3$ that form a partition of $[n]$.  With $V=[6]$ an example of this type saturating the bound is
$\Gcal = \{ \{1\}, \ldots, \{6\}, \{1,2\}, \{3,4\}, \{5,6\} \}$.  The second is the case where there are two maximal sets $X_1,X_2$ that form a
partition of $[n]$ and exactly one of $X_1,X_2$ is not included.  The latter includes the case where there is a single maximal set $X_1$ of size $|X_1| = n-1$.
For $V=[6]$, an example of this type is $\Gcal = \{\{2\}, \ldots, \{6\}, \{2,3\},\{2,3,4\},\{2,3,4,5\},\{2,3,4,5,6\}\}$.
\end{remark}

Chandran and Ram (Lemma~2.13 in \cite{CR04}) show that if the set $\Mcal(G)$ of minimum cuts of a
graph $G$ is cross-free, then $|\Mcal (G)| \le 2n-3$.  This is an
easy corollary of \cref{lem:laminar_size}, which gives something more general.

\begin{corollary}
\label{cor:cross-free_size}
Let $G = (V,w)$ be a graph on $n \ge 2$ vertices.
Let $\Lcal \subseteq \Mcal(G)$ be a subset of minimum cuts that is cross-free.  Then
$|\Lcal| \le 2n-3$.
\end{corollary}

\subsection{Spanning}
Let $\Lcal \subseteq \Mcal(G)$ be a maximal cross-free subset of $\M(G)$.  Here maximal means that for any cut $S \in \M(G) \setminus \Lcal$ there
is a cut $T \in \Lcal$ that crosses $S$.  The fact that $\laspan(\vec{\Lcal}) = \laspan(\vec{\Mcal}(G))$
essentially follows from a key lemma of Jain in his factor of 2 approximation algorithm for the survivable network design problem (Lemma~4.2 in \cite{Jain01}).
Another application of a similar lemma can be found in Goeman's approximation algorithm for the bounded-degree minimum spanning tree problem \cite{Goemans06}.

The context of Jain's lemma is slightly different than ours, as we now explain.  Instead of mincuts, Jain considers the set of cuts
$\Tcal$ which saturate the inequalities of a particular linear program.  He shows that the set $\Tcal$ has the property that if $\Delta(X),\Delta(Y) \in \Tcal$
cross then either
\begin{enumerate}
\item $\Delta(X \cap Y), \Delta(X \cup Y) \in \Tcal$ and $\chi(\Delta(X)) + \chi(\Delta(Y)) = \chi(\Delta(X \cap Y)) + \chi(\Delta(X \cup Y))$, or
\item $X \setminus Y, Y \setminus X \in \Tcal$ and $\chi(\Delta(X)) + \chi(\Delta(Y)) = \chi(\Delta(X \setminus Y)) + \chi(\Delta(Y \setminus X)).$
\end{enumerate}
As shown by Dinitz, Karzanov, and Lomonosov \cite{DKL76}, for crossing mincuts $\Delta(X),\Delta(Y)$ \emph{both} items (1), (2) hold (see \cref{prop:bixby} for a proof).
Thus Jain's lemma applies to $\M(G)$ as well.
\begin{restatable}[\cite{Jain01}]{lemma}{jainspan}\label{thm:span}
Let $G = (V,w)$ be a graph and $\Lcal \subseteq \M(G)$ be a maximal cross-free family of mincuts.
Then $\laspan(\vec{\Lcal}) = \laspan(\vec{\M}(G))$.
\end{restatable}
For completeness, we include a full proof of \cref{thm:span} in \cref{sec:appendix}.

We now can give the first proof of our main upper bound that for any $n\ge 2$ an $n$-vertex graph $G=(V,w)$ has $\cd(G) \le 2n-3$.

\begin{proof}[Proof of \cref{thm:cd_upper}]
Follows from \cref{cor:cross-free_size} and \cref{thm:span}.
\end{proof}


\section{Explicit construction of graphs with cut dimension $2n-3$}\label{sec:construction}
In this section we prove \cref{thm:cd_lower} by giving a general technique to explicitly construct graphs of cut dimension $2n-3$.  We focus
on constructing graphs $G = (V,w)$ where $w$ is strictly positive, i.e.\ where $G$ is a complete weighted graph.
The main lemma of this section, \cref{prop:lin_ind}, shows that, in a complete weighted graph, for any cross-free family of cuts $\Lcal$
the vectors in $\vec{\Lcal}$ are linearly independent.

Thus to construct a graph with cut dimension $2n-3$ it suffices to construct a complete weighted graph whose set of mincuts
is a cross-free family of cuts of cardinality $2n-3$.  Such a graph is constructed for every $n \ge 2$ in Theorem~5.2 of \cite{CR04}.
Combining this construction with our linear independence result \cref{prop:lin_ind} gives a proof of our main lower bound \cref{thm:cd_lower}.

In \cref{subsec:construction} we go further and show for any maximal cross-free family $\Fcal \subseteq 2^{[n]}$ there is a complete weighted graph $G = ([n],w)$
with $\M(G) = \{ \Delta(X) : X \in \Fcal\}$.  Moreover, we give an explicit formula for the weight vector $w$.
Part of this construction is a lemma, \cref{lem:min_value},
which may be of independent interest: it says that if $\Lcal$ is a maximal family of cross-free cuts in a graph $G$, and all cuts in $\Lcal$ have the same weight $c$, then $c$
is the weight of the minimum cut in $G$.

A key tool for showing the linear independence of cuts from a cross-free family is the tree representation of a laminar family, which we go over next.

\subsection{Tree representation}
\begin{definition}
For an unweighted directed graph $G=(V,E)$ we let $\delta^+(X) = \{(x,y) \in E : x \in X, y \in V - X\}$.  For a singleton $v \in V$
we write $\delta^+(v)$ instead of $\delta^+(\{v\})$.
\end{definition}

\begin{definition}[Arborescence]
An \emph{arborescence} is a directed rooted tree where all edges point away from the root.  A vertex of an arborescence which is
not the root or a leaf we call an \emph{internal vertex}.
\end{definition}

\begin{definition}[Tree representation]\label{def:tree-repre}
Let $T$ be a directed graph whose underlying undirected graph is a tree.  Let $U$ be a finite set and $\phi: U \rightarrow V(T)$.
For $e = (x,y) \in E(T)$ define $S_e$ as
\[
S_e = \{ s \in U : \phi(s) \mbox{ is in the same connected component of } T - e \mbox{ as } y \} \enspace .
\]
Then $(T,\phi)$ defines a set family $\Fcal = \Fcal(T, \phi)$ where $\Fcal = \{S_e : e \in E(T)\}$.
We say that $(T,\phi)$ is a \emph{tree representation} of $(U, \Fcal)$.
We call $(T,\phi)$ a \emph{faithful tree representation} if $|E(T)| = |\Fcal|$.
For $v \in V(T)$, if there is a $u \in U$ such that $\phi(u) = v$ then we say that $v$ has a \emph{label}.
\end{definition}

We will need the fact that a laminar set family has a faithful tree representation by an arborescence.  A textbook proof of this fact can be found
in Korte and Vygen Proposition 2.14 \cite{KV18}.  While they do not explicitly say the tree representation they construct is faithful,
this is clear from the proof.
\begin{proposition}
\label{prop:treerep}
Let $(U,\Fcal)$ be laminar family.  Then there is a faithful tree representation $(T,\phi)$ of $(U,\Fcal)$ where $T$ is an arborescence.
\end{proposition}

Recall from \cref{claim:cross-free} that if $\Lcal$ is a cross-free family of cuts then its beach $\Gcal$ is laminar, and thus has a tree representation.
\begin{lemma}[Tree structure of maximal cross-free families]
\label{lem:tree_structure}
Let $\Lcal$ be a maximal family of cross-free cuts of a graph $G= ([n],w)$ and $\Gcal \subseteq 2^{[n]}$ its beach.  Then in a faithful tree representation $(T,\phi)$ of $\Gcal$
it holds that
\begin{enumerate}
\item The root $r$ is labeled by $1$ and has $|\delta^+(r)| = 1$
\item There are $n-1$ leaves of $T$ each with a distinct label in $\{2, \ldots, n\}$.
\item Every internal vertex $v$ has $|\delta^+(v)| = 2$.
\end{enumerate}
\end{lemma}

\begin{proof}
As by the definition of a beach, sets do not contain $1$, this means that $1$ must be the label of the root.
As star cuts do not cross any other cut, if $\Lcal$ is maximal it must contain all the star cuts.  This means
that $\Gcal$ contains the sets $\{2\}, \ldots, \{n\}, \{2,\ldots, n\}$.  Thus the outdegree of the root must be $1$, as
this outgoing edge represents the set $\{2,\ldots, n\}$.  Further there must be $n-1$ leaves which are labeled by $2, \ldots, n$.
We have now accounted for all the labels, thus no internal vertex has a label. Further, if there was a leaf $v$ with parent $u$
such that $v$ did not have a label, then $(u,v)$ would represent the
empty set, which by definition is not in $\Gcal$.  Thus there are exactly $n-1$ leaves.

It remains to show that every internal vertex $v$ of $T$ which is not the root has $|\delta^+(v)| = 2$.  Let $v$ be an internal vertex, and as $v$ is not
the root, let $u$ be its parent, and as $v$ is not a leaf let $w$ be a child of $v$.  If $|\delta^+(v)| = 1$ then the edges $(u,v),(v,w)$ would represent the same set,
as $v$ is not labeled.  This contradicts the fact that $(T,\phi)$ is a faithful tree representation.  Now suppose $|\delta^+(v)| > 2$ and let $w,x,y$ be three of its children.  Consider
the sets $X_1, X_2,X_3 \in \Gcal$ represented by the edges $(v,w),(v,x),(v,y)$.  Further the edge $(u,v)$ represents a set $A \in \Gcal$ with
$X_1 \cup X_2 \cup X_3 \subseteq A$.
We claim that in this case $\Lcal$ is not maximal because the cut $\Delta(X_1 \cup X_2)$ does not cross any cut in $\Lcal$.  Indeed, $X_1 \cup X_2$ is contained
in all the sets represented by edges on the path from $v$ to the root, and is disjoint from the sets represented by any other edge of $T$.  Thus we have a contradiction.
\end{proof}

\begin{corollary}
\label{cor:maxcard}
Let $\Lcal$ be a maximal family of cross-free cuts of a graph $G= ([n],w)$.  Then $|\Lcal| = 2n-3$.
\end{corollary}

\begin{proof}
Let $\Gcal$ be the beach of $\Lcal$ and $(T,\phi)$ a faithful tree representation of $\Gcal$.  As $(T,\phi)$
is faithful $|E(T)| = |\Lcal|$.  Let $T'$ be the undirected graph underlying $T$.  Clearly $|E(T')| = |E(T)|$.  We use \cref{lem:tree_structure} to count $|E(T')|$.
Let $i$ be the number of internal vertices of $T'$, each of which has degree 3.  There are also $n$ non-internal vertices each of which has
degree $1$.  Thus $|E(T')| = (3i + n)/2$.  Also as $T'$ is a tree $|E(T')| = |V(T')|-1 = n+i-1$.  Hence $i = n-2$ and $|\Lcal| = |E(T)| = |E(T')| = 2n-3$.
\end{proof}

\subsection{Linear independence}
We now show the main theorem of this section that in a complete weighted graph any set $\vec{\Lcal}$ of cut vectors of a cross-free family of cuts $\Lcal$ is linearly independent.
We will use the tree representation $(T,\phi)$ of the beach $\Gcal$ of $\Lcal$ to do this via the following lemma.
\begin{lemma}
\label{lem:tree_assign}
Let $T$ be an arborescence with root $r$ and $\psi : E(T) \rightarrow \R$.   Let $U$ be a finite set and $\phi: U \rightarrow V(T)$.  Suppose that $T,\phi, \psi$ have the property
that
\begin{enumerate}
  \item The root $r$ is labeled and has $|\delta^+(r)| = 1$.
  \item Every internal vertex $v$ is unlabeled and has $|\delta^+(v)| = 2$.
  \item Every leaf of $T$ has a label.
  \item For every $s,t \in U$ it holds that $\sum_{e \in \phi(s)-\phi(t)} \psi(e) = 0$, where $\phi(s)-\phi(t)$ is the set of edges on the undirected path from $\phi(s)$ to $\phi(t)$.
\end{enumerate}
Then $\psi$ is identically $0$.
\end{lemma}

\begin{proof}
We will prove by induction on the depth of the arborescence.  We need a slightly different statement for the inductive hypothesis
since when considering a sub-arborescence $T'$ of $T$ we do not know that the root of $T'$ has property (1).

\paragraph{Inductive hypothesis:} Let $T$ be an arborescence with root $r$ that is unlabeled and has $|\delta^+(r)| = 2$, and further suppose
$T, \phi, \psi$ satisfy conditions (2)-(4) of the proposition.  Then letting $u,v$ be the children of $r$ it holds that
$\psi((r,u)) = -\psi((r,v))$ and for any other edge $e \in E(T), e \ne (r,u),(r,v)$ it holds that $\psi(e) = 0$.

For the base case consider a tree of depth $1$, with root $r$ and two children $u,v$ which are leaves.  As they are leaves, $u,v$ are
labeled which, considering the path from $u$ to $v$, means $\psi((r,u)) + \psi((r,v)) = 0$.  This concludes the base case.

Now we prove the inductive step.  Let $r$ be the root of a tree with children $u,v$.  We consider two cases:
\paragraph{Case 1: one of $u,v$ is a leaf.} Suppose without loss of generality that $u$ is a leaf and $v$ is an internal node with children $v_1, v_2$.
By the inductive hypothesis $\psi((v,v_1)) + \psi((v,v_2)) = 0$ and $\psi$ is identically $0$ on the subtrees rooted at $v_1,v_2$.  Let $y_1, y_2$ be leaves
that are descendants of $v_1,v_2$ respectively (and can possibly be $y_1,y_2$ themselves).  Considering the path from $u$ to $y_1$ and $y_2$ we have
the equations
\begin{align*}
\psi((r,u) + \psi((r,v)) + \psi((v,v_1)) &= 0 \\
\psi((r,u) + \psi((r,v)) + \psi((v,v_2)) &= 0
\end{align*}
As $\psi((v,v_1)) + \psi((v,v_2)) = 0$, adding these equations shows that $\psi((r,u) + \psi((r,v)) =0$, as desired.  Substituting this back
into the equations further implies that $\psi((v,v_1)) = \psi((v,v_2)) = 0$ so $\psi$ is identically $0$ on the subtree rooted at $v$ completing
this case.

\paragraph{Case 2: both $u,v$ are internal vertices.} Let the children of $u$ be $u_1, u_2$ and the
children of $v$ be $v_1, v_2$.
By the inductive hypothesis, $\psi(\cdot)$ is identically zero on the sub-trees rooted at $u_1, u_2, v_1, v_2$
and we have $\psi((u,u_1)) + \psi((u,u_2)) = \psi((v,v_1)) + \psi((v,v_2))= 0$.  We must show that $\psi((u,u_1))= \psi((u,u_2))=\psi((v,v_1)) =\psi((v,v_2))= 0$
and that $\psi((r,u)) + \psi((r,v)) = 0$.

Let $x_1,x_2$ be a leaves that are descendants of $u_1,u_2$, respectively, and similarly let $y_1, y_2$ be leaves that are
descendants of $v_1,v_2$, respectively.  By assumption all of these leaves are labeled.  Considering the paths from $x_b-y_{b'}$ for $b,b' \in \{0,1\}$ we obtain
the following four constraints on $\psi$:
\begin{align*}
\psi((u,u_1)) + \psi((r,u)) + \psi((r,v)) + \psi((v,v_1)) &= 0 \\
\psi((u,u_1)) + \psi((r,u)) + \psi((r,v)) + \psi((v,v_2)) &= 0 \\
\psi((u,u_2)) + \psi((r,u)) + \psi((r,v)) + \psi((v,v_1)) &= 0 \\
\psi((u,u_2)) + \psi((r,u)) + \psi((r,v)) + \psi((v,v_2)) &= 0 \\
\end{align*}
Adding all four equations and using $\psi((u,u_1)) + \psi((u,u_2)) = \psi((v,v_1)) + \psi((v,v_2))= 0$ shows that $\psi((r,u)) + \psi((r,v)) = 0$.
Taking this into account, adding the first two equations then shows $\psi((u,u_1)) = 0$, and adding the last two equations shows $\psi((u,u_2)) = 0$.
This then also means $\psi((v,v_1)) = \psi((v,v_2)) = 0$.

We have now shown the inductive statement holds.  It remains to see why this implies the lemma.  Let $r$ be the root of the tree, let $u$ be the
child of $r$, and let $u_1, u_2$ be the children of $u$.  By the inductive statement we have that $\psi((u,u_1)) + \psi((u,u_2)) = 0$ and $\psi$ is
identically zero on the subtree rooted at $u_1$ and the subtree rooted at $u_2$.  Let $x_1,x_2$ be leaves which are descendants of $u_1, u_2$, respectively.
As the root has a label, considering the path from $r$ to $u_1$ implies that $\psi((r,u)) + \psi((u,u_1)) = 0$ and considering the path from $r$ to $u_2$
implies $\psi((r,u)) + \psi((u,u_2)) = 0$.  Adding these equations implies that $\psi((r,u)) = 0$, from which it then follows that $\psi((u,u_1)) = \psi((u,u_2)) = 0$.
\end{proof}

\begin{lemma}
\label{prop:lin_ind}
Let $G = ([n],w)$ be a complete weighted graph and let $\Lcal$ be a cross-free family of cuts.  Then $\vec{\Lcal} = \{\chi(S): S \in \Lcal\}$ form a linearly independent set of vectors.
\end{lemma}

\begin{proof}
We may assume that $\Lcal$ is a \emph{maximal} cross-free family, as showing that a superset of $\vec{\Lcal}$ is linearly independent
implies that $\vec{\Lcal}$ is as well.  Thus suppose $\Lcal$ is a maximal cross-free family and let $\Gcal$ be its beach.  Let $(T,\phi)$ be a faithful tree representation of
$\Gcal$.  By \cref{lem:tree_structure} we have that $(T,\phi)$ satisfy conditions (1)-(3) of \cref{lem:tree_assign}.

Now we ask the question: for an edge $\{i,j\} \in E(G)$ which sets $S \in \Lcal$ contain it?
This has a very nice description in terms of the tree decomposition.  Let $u,v \in V(T)$ be the vertices with
$\phi(i) = u, \phi(j) = v$.  Then the sets containing $i$ are the sets represented by edges from the root to $u$;  the sets containing
$j$ are the sets represented by the edges on the path from the root to $v$.  Therefore the sets which contain $i$ but not $j$ or $j$ but
not $i$, are exactly those represented by the edges on the path from $u$ to $v$ in the undirected tree underlying $T$.  Thus the cuts which
contain the edge $\{i,j\}$ are exactly those with a shore which is represented by an edge on the path from $u$ to $v$ in undirected graph underlying $T$.

Consider a linear combination $\sum_{S \in \Lcal} \alpha_S \chi(S) = \zero$ which is equal to the all zero vector.  The $\{i,j\}$ coordinate
of this equation says that $\sum_{S \in \Lcal, \{i,j\} \in S} \alpha_S \chi(S)(\{i,j\}) = 0$.  This sum is exactly over the sets represented by edges
on the path from $\phi(i)$ to $\phi(j)$.  As this sum must be zero for every edge $\{i,j\}$, this says that if we let $\psi(e) = \alpha_S$ where the edge $e$
represents a shore of $S$ then for any two labeled vertices $u,v \in V(T)$ the sum of $\psi(e)$ over the edges on the path from $u$ to $v$ is zero.
Thus also condition~(4) of \cref{lem:tree_assign} is satisfied.
Hence all of the conditions of \cref{lem:tree_assign} hold which implies that
$\psi$ must be identically zero and therefore all coefficients $\alpha_S = 0$.  This shows that $\{\chi(S): S \in \Lcal\}$ is a linearly independent set.
\end{proof}

We can now give the first proof of our main lower bound result on the cut dimension \cref{thm:cd_lower}, which
says that for every integer $n\ge 2$ there is an $n$-vertex weighted graph $G = (V,w)$ with $\cd(G) \ge 2n-3$.

\begin{proof}[Proof of \cref{thm:cd_lower}]
For every integer $n \ge 2$, Theorem~5.2 of \cite{CR04} constructs a complete weighted graph $G = (V,w)$ on $n$ vertices such that
$\M(G)$ is a cross-free family of size $|\M(G)| = 2n-3$.  By \cref{prop:lin_ind} the vectors in $\vec{\Mcal}$ form a linearly
independent set, thus $\cd(G) \ge 2n-3$.
\end{proof}

\subsection{Constructing graphs with a cross-free set of mincuts}
\label{subsec:construction}
In this subsection we explicitly construct, for any maximal cross-free family $\Fcal \subseteq 2^{[n]}$, a complete weighted graph $G = ([n],w)$
with $\M(G) = \{ \Delta(X) : X \in \Fcal\}$.  This task is made easier by the next lemma.  We first need a definition.

\begin{definition}
\label{def:overlap}
Let $\Fcal \subseteq 2^{V}$.  For a subset $X \subseteq V$, let $\cross_{\Fcal}(X) = \{Y \in \Fcal: X, Y \ \cross\}$.
\end{definition}

\begin{lemma}
\label{lem:min_value}
Let $G = (V,w)$ be a graph and $\Lcal$ be a maximal cross-free family of cuts.  Suppose that for all
$S \in \Lcal$ it holds that $w(S) = c$.  Then the weight of a minimum cut in $G$ is $c$.
\end{lemma}

\begin{proof}
Let $\Gcal$ be the beach of $\Lcal$.  Suppose for a contradiction that the weight of a minimum cut of $G$ is $< c$.
Let $\Tcal = \{ Z : \emptyset \ne Z \subsetneq V, v_1 \not \in Z, Z \not \in \Gcal, w(\Delta(Z)) < c\}$ and
\[
X = \argmin_{Z} \{|\cross_\Gcal(Z)|: Z \in \Tcal \} \enspace .
\]
In the following we always use $\cross(\cdot)$ with respect to $\Gcal$ and
drop the subscript.  As $|\cross(X)| \ge 1$, let $Y \in \cross(X)$.  As shown in \cref{sec:appendix} \cref{lem:intersect}, both
$|\cross(X \cap Y)|$ and $|\cross(X \cup Y)|$ are strictly smaller than $|\cross(X)|$.  Thus it must be the case that
$X \cap Y, X \cup Y \not \in \Tcal$.  Let us take the case of $X \cap Y$.  It does not contain $v_1$, as neither $X$ nor $Y$ do,
and it is a nonempty set by the definition of $\cross$.  Thus it must be the case that either $w(\Delta(X \cap Y)) \ge c$ or
that $X \cap Y \in \Gcal$, which implies $w(\Delta(X \cap Y)) = c$.  The same argument holds for $X \cup Y$, thus
both $w(\Delta(X \cap Y)), w(\Delta(X \cup Y)) \ge c$.

However by submodularity of the cut function we have
$w(\Delta(X\cap Y )) + w(\Delta(X \cup Y)) \le w(\Delta(X)) + w(\Delta(Y))$, which implies that at least one of $\Delta(X \cap Y), \Delta(X \cup Y)$ must have weight $<c$.
Hence we have a contradiction and the lemma holds.
\end{proof}

We will additionally need the following theorem which follows from Theorem 5.1 in \cite{CR04}.
\begin{theorem}[\cite{CR04}]
\label{thm:cr_complete}
Let $G = (V,w)$ be a complete weighted graph.  Then $\M(G)$ is a cross-free family of cuts.
\end{theorem}

\begin{theorem}
\label{thm:explicit}
Let $n \ge 2$ and $\Lcal$ be a {maximal} cross-free family of cuts in the $n$-vertex complete graph.  Let
$A$ be an $|\Lcal|$-by-$\binom{n}{2}$ matrix whose rows are the vectors $\chi(S)$ for $S \in \Lcal$ and
let $z = A^T \1$.  Define $w(e) = 2^{-z(e)+1}$ for $e \in [n]^{(2)}$.
Then $G=([n],w)$ is a complete weighted graph with $\cd(G) = 2n-3$ and $\M(G) = \Lcal$.
\end{theorem}

\begin{proof}
It is clear from the definition that $w >0$ and so defines a complete weighted graph.
We will show that $Aw = \1$.  By \cref{lem:min_value} this shows that
the minimum cut weight of $G$ is $1$ and so the set of minimum cuts includes $\Lcal$.  As $w$ defines a complete weighted graph, by \cref{thm:cr_complete}
the set of minimum cuts in $G$ is cross-free and therefore must be exactly $\M(G) = \Lcal$, since $\Lcal$ is maximal.  Further, {$|\Lcal| = 2n-3$ by \cref{cor:maxcard} and}
the vectors in $\vec{\Lcal}$ are linearly independent by \cref{prop:lin_ind}, thus $\cd(G) = 2n-3$.

It remains to show $Aw = \1$.
We do this using an alternative way of viewing the assignment of edge weights.
Let $\Gcal \subseteq 2^{[n]}$ be the beach of $\Lcal$, and $(T,\phi)$ be a faithful
tree representation of $\Gcal$.  For vertices $u,v \in V(T)$ let $d(u,v)$ be the length of the shortest path between $u,v$ in the undirected graph underlying $T$.
Now let $\{i,j\} \in [n]^{(2)}$ and suppose $\phi(i) = u, \phi(j) = v$.  We claim that $w(\{i,j\}) = 2^{-d(u,v)+1}$.
The sets of $\Gcal$ containing $i$ are the sets represented by edges from the root to $u$;  the sets of $\Gcal$ containing
$j$ are the sets represented by the edges on the path from the root to $v$.  Therefore the sets which contain $i$ but not $j$ or $j$ but
not $i$, are exactly those represented by the edges on the path from $u$ to $v$ in the undirected tree underlying $T$.  As $(T,\phi)$ is faithful,
each of these edges represents a different set, and therefore the number of edges on the path from $u$ to $v$ is exactly the number of
sets of $\Lcal$ which contain $\{i,j\}$.

We now continue with the proof that $Aw = \1$ using this interpretation of the weights.  For any cut $S\in \Lcal$ with shore $X \in \Gcal$, take the edge $(u,v) \in E(T)$ representing $X$.
Now imagine we remove the edge $(u,v)$ from $T$ which disconnects $T$ into two components.  Let $T_u$ be the component containing $u$ and $T_v$ the component containing $v$.
From $T_u$, which contains the root $r$ of $T$, we create a graph $T_u'$ whose underlying undirected graph is the same as $T_u$, but for which all edges are directed away from $u$.  Thus in
$T_u'$, vertex $u$ becomes the root and $r$ becomes a leaf.  Now by item~(2) of \cref{lem:tree_structure}, every non-leaf vertex in $T_v$ and $T_u'$ has out-degree $2$.  We inject a unit of flow into $u$ in the graph $T_u'$
and let it propagate according to the rule that at every non-leaf vertex half of the flow is routed along each outgoing edge.  We similarly inject a unit of flow into $v$ in the graph $T_v$ and let it
propagate according to the same rule.  Thus in the tree $T_v$, each leaf $a$ gets $f(a) = 2^{-d(a,v)}$ amount of flow, where $d(a,v)$ is the number of edges along the path from $v$ to $a$ in $T_v$.
Similarly, if $b$ is a leaf in the tree $T_u'$, the amount of flow arriving at $b$ is $f(b) = 2^{-d(b,u)}$. Now let $\{i,j\} \in [n]^{(2)}$ with $i \in X, j \in \bar{X}$ and observe that the way we defined $w(\{i,j\})$ satisfies
 \[
 w(\{i,j\}) = 2^{-d(\phi(i),\phi(j))+1}  = 2^{-d(\phi(i),v) - d(\phi(j),u)} = f(\phi(i))\cdot  f(\phi(j)) \enspace.
 \]
 Thus the weight of the cut $S$ is
 \[
 \sum_{i\in X,j\in \bar{X}} w(\{i,j\}) =  \sum_{i\in X,j\in \bar{X}} f(\phi(i)) \cdot f(\phi(j)) =  \left(\sum_{i\in X} f(\phi(i))\right) \cdot \left(\sum_{j\in \bar{X}} f(\phi(j))\right) = 1\cdot 1 = 1 \enspace .
 \]

\end{proof}


\section{Another proof using graph operations}\label{sec:another-proof}
In this section we give another proof of our main theorems: we prove that the cut dimension of any $n$-vertex graph is at most $2n-3$
and we also prove that this upper bound is tight.
An important role will be played by the following lemma, giving an explicit characterization of graphs having at least one non-star mincut, where none of these mincuts is crossless. This characterization has
originally appeared in \cite{bixby75,DKL76}.  More modern presentations can be
found in Lemma~2.9 of \cite{CR04} or Lemma~2 of \cite{FF09}.
\begin{lemma}
\label{lem:crossing_cycle}
Suppose that $G=(V,w)$ is a graph which has a non-star mincut, and every non-star mincut is crossed by a non-star mincut.
Then $G$ is a cycle where all edges have the same weight.
\end{lemma}

Let us denote by $C_n$ the cycle on the $n$ vertex set
$V=\{v_1, \ldots, v_n\}$ and with edge set $E = \{ \{v_1, v_2\}, \ldots , \{v_{n-1}, v_n\}, \{v_n , v_1\}\}$,
where the weight of every edge is the same.
We also need that the cut dimension of $C_n$ is at most $n$. In fact, it is easy
to prove that the its cut dimension is exactly $n$ when $n\geq 3$.
\begin{lemma}
\label{lem:cycledim}
The cut dimension of $C_2$ is $1$, and $\cd(C_n) = n$, for $n \geq 3$.
\end{lemma}

\begin{proof}
The statement for $n=2$ is obvious.  For $n \ge 3$ we have $\cd(C_n) \le n$ as the graph only has $n$ edges
and thus the cut vectors are elements of $\R^n$ which has dimension $n$.

For the lower bound we construct a set of $n$ linearly independent minimum cut vectors in $C_n$.
Label the coordinates of the vectors by the edges $\{v_1, v_2\}, \ldots , \{v_{n-1}, v_n\}, \{v_n , v_1\}$.
We define the sets $X_1 =  \{v_1, v_2\}$ and $X_k = \{v_2, \ldots, v_k\}$, for $2 \leq k \leq n$.

We claim that the cut vectors $\xi_k = \chi(\Delta(X_k))$, for $1 \leq k \leq n$, are linearly independent.
Let $e_i$ be the $\ith$ standard basis vector in $\R^n$.  Then we see that
$\xi_1 = e_2  + e_n$ and $\xi_k = e_1 + e_k$, for $2 \leq k \leq n$.
Thus $\xi_2 + \xi_n - \xi_1 = 2e_1$, so $e_1$ is in the span of these vectors.  Also
$e_k = \xi_k - e_1$ is in the span for $2 \leq k \leq n$. Hence these $n$
vectors span all of $\R^n$ and therefore must be linearly independent.
\end{proof}

\subsection{Two lemmas on graph operations}

The main technical part of the second proof of our main theorems is played by the two lemmas in this section.
The second lemma gives an upper bound on the cut dimension of a graph $G$  in function of the cut dimension
of the smaller graphs obtained when $G$ is separated along a crossless non-star minimum cut $Z$.
Moreover, this upper bound becomes an equality when in addition the cut $Z$ is connected. Our upper and lower bounds
for the cut dimension are respectively almost immediate consequences of these results.
\begin{lemma}
\label{lem:crosslessA}
Let $G=(V,w)$ be a weighted graph and let $Z \in \M(G)$ be a crossless non-star minimum cut defined by shores $X_0, X_1 = V \setminus X_0$.
For $b \in \{0,1\}$, let ${\M}_{b} = \{S \in \M(G) : S \subseteq Z \cup E(X_{b}) \}$.
Let $\sep(G,Z) = \{ G_0= (V_0, w_0), G_1 = (V_1,w_1) \}$ as defined in \cref{sec:prelim}, where $V_b = X_b \cup \{v_{1-b}\}$, for
$b \in \{0,1\}$, with $v_0, v_1 \not \in X_0 \cup X_1$. Then $\dim(\laspan(\vec{\M}_b)) = \cd(G_b)$, for $b \in \{0,1\}$.
\end{lemma}

\begin{proof}
We prove the statement for $b=0$, the other case follows in exactly the same manner.
Let $m = |E|$ and partition $E$ into three disjoint sets $E=E(X_0) \sqcup Z \sqcup E(X_1)$.
Call a vertex $x \in X_0$ \emph{friendly} if it has a neighbor in $X_1$, that is
there exists an edge $\{x,y\} \in Z$ for some $y \in X_1$.
The edges in $Z$  can then be partitioned into the disjoint union of sets $Z_x$, over all friendly $x$,
where $Z_x = \{ e \in Z: x \in e\}$.

Let $\M(G_0)$ be the set of all minimum cuts of $G_0$.
The set $\vec{\M}(G_0)$ is composed of $m_0$ dimensional vectors
where $m_0 = |E(X_0) | + \deg(v_1)$.
Observe that $\deg(v_1)$ is the number of friendly vertices in $X_0$.
We can partition the edges of $G_0$ into two sets
$E(X_0) \sqcup Z_1$ where $Z_1 = \{ \{x, v_1\}: x  {\rm ~ is ~ friendly } \}$.

We define a natural bijection
$\psi : {\M}_{0} \rightarrow {\M}(G_0)$ as follows. Let
$S$ be a mincut in ${\cal M}_{0}$ with shores $X'$ and $V \setminus X'$, where $X' \subseteq X_0$. Note that we can assume this because $Z$ is crossless.
Then $\psi(S)$ is the mincut in $\M(G_0)$ whose shores are $X'$ and
$( X_0 \setminus X' ) \cup \{v_1\}$.  Let $k = |\M_{0}| = | \M(G_0)|$.

We now consider two matrices $C$ and $D$, where $C$ is a $k$-by-$m$ matrix and $D$ is a $k$-by-$m_0$ matrix.
Fix an ordering $S_1, \ldots, S_k$ of $\M_0$ and let the  $\ith$
row of $C$ be $ \chi(S_i)$, the characteristic vector of the cut $S_i$.
Likewise the $\ith$ row of
$D$ is $\chi(\psi(S_i))$.
We have $\rk(C) = \dim ( \laspan ( \vec{\M}_{0}))$ and $\rk(D)= \cd(G_0)$.

The columns of $C,D$ are labeled by edges.  For $C$, we label the edges according to
the partition $E=E(X_0) \sqcup Z \sqcup E(X_1)$, with edges in $E(X_0)$ coming first, then edges
from $Z$, then edges from $E(X_1)$.  For $D$, we label the edges according to the
partition $E(X_0) \sqcup Z_1$, again with edges from $E(X_0)$ coming first and then those from $Z_1$.
We observe the following facts:
\begin{itemize}
\item The edges in $E(X_0)$ are common in $G$ and $G_1$, and $ \chi(\psi(S_i))(e)  = \chi(S_i)(e)$,
for every $S_i \in \M_0$ and edge $e \in E(X_0)$. This means that columns of $C$ and $D$ labeled
by an edge $e \in E(X_0)$ are identical.

\item For an edge $e \in E(X_1)$, we have that $\chi(S_i)(e)=0$, for every $S_i \in \M_0$.  Thus
columns of $C$ labeled by an edge $e \in E(X_1)$ are all zero.

\item Finally, for a friendly $x \in X_0$ consider any edge $e = \{x,y\} \in Z_x$ and the edge
$f = \{x,v_1\} \in Z_1$. Then the $e^{\scriptsize \mbox{{\rm th}}}$
column of $C$ and the $f^{\scriptsize \mbox{{\rm th}}}$ column of $D$ are identical
because for every $S_i \in \M_0$
we have $\chi(S_i)(e) = 1$ iff $x \in X'$  iff $\chi(\psi(S_i))(f) = 1$.
\end{itemize}
These points together imply that $D$ is actually a submatrix of $C$, which can be obtained by taking the columns labeled by edges in $E(X_0)$
and then taking $|Z_1|$ more columns of $C$ by choosing one $e \in Z_x$ for every friendly $x \in X_0$. Therefore $\rk(D) \le \rk(C)$.

We can also see that $\rk(C) \le \rk(D)$ as $C$ can be obtained from $D$ by repeating columns labeled by edges in
$Z_1$ several times and adding all zero columns, and neither of these operations increase the rank.
\end{proof}

\begin{lemma}
\label{lem:crosslessB}
Let $G, Z, G_0, G_1$ as in~\cref{lem:crosslessA}.
Then $\cd(G) \leq \cd(G_0) + \cd(G_1) - 1$, and if $Z$ is connected then the equality holds.
\end{lemma}

\begin{proof}
We first prove that $\cd(G) \leq \cd(G_0) + \cd(G_1) - 1$.
The important fact is that $\M(G) \subseteq \M_{0} \cup \M_1$ because $Z$ is a crossless mincut.  Also since $\M_0, \M_1 \subseteq \M(G)$ we in fact have
$\M(G) = \M_{0} \cup \M_1$.  Therefore
\begin{align*}
\cd(G) &= \dim(\laspan(\vec{\M}(G))) \\
&= \dim ( \laspan ( \vec{\M}_{0} \cup \vec{\M}_{1})) \\
&= \dim ( \laspan ( \laspan (\vec{\M}_{0} ) \cup  \laspan (\vec{\M}_{1} ))) \\
&= \dim ( \laspan ( \vec{\M}_{0})) + \dim ( \laspan ( \vec{\M}_{1})) -
\dim ( \laspan ( \vec{\M}_{0}) \cap  \laspan ( \vec{\M}_{1})) \\
&=  \cd(G_0) + \cd(G_1) - \dim ( \laspan ( \vec{\M}_{0}) \cap  \laspan ( \vec{\M}_{1})) \enspace .
\end{align*}

We use
\cref{lem:crosslessA}  to obtain the last equality.
Notice that $Z \in \M_0 \cap \M_1$, which implies that $\dim ( \laspan ( \vec{\M}_{0}) \cap  \laspan ( \vec{\M}_{1}))\geq 1$, and thus
$\cd(G) \leq \cd(G_0) + \cd(G_1) - 1$.

We now prove the inequality in the reverse direction, when $Z$ is connected.
Let $d_b = \cd (G_b)-1$, for $b=0,1$.
Let $Z_b$ be the star cut  at $v_{1-b}$ in $G_b$. Since these are mincuts, we can extend them to a basis
in the respective graphs.
Therefore there exist $A_1, \ldots A_{d_0} \subset X_0$ and $B_1, \ldots B_{d_1} \subset X_1$ such that the family
$\{ \chi(\Delta  (A_1)), \ldots, \chi(\Delta  (A_{d_0})) , \chi(Z _0) \}$ is independent in
$\laspan(\vec{\M}(G_0))$
and the family $\{ \chi(\Delta  (B_1)), \ldots, \chi(\Delta  (B_{d_1})) , \chi(Z _1)\}$ is independent in
$\laspan(\vec{\M}(G_1)$.
We claim that in $\laspan(\vec{\M}(G))$ the set
$\{ \chi(\Delta  (A_1)), \ldots, \chi(\Delta  (A_{d_0})) , \chi(\Delta  (B_1)), \ldots, \chi(\Delta  (B_{d_1})) , \chi(Z )\}$
of size $d_0 + d_1 +1$ is independent.

Let us suppose on the contrary that a non-trivial linear combination of these $d_0 + d_1 +1$
vectors gives $\zero$.
Then there exist non all zero real numbers
$a_1, \ldots, a_{d_0}, b_1, \ldots , b_{d_1}$ and $\varepsilon \in \{0,1\}$
such that
\begin{equation}
\label{eq:dependent}
\sum_{i=1}^{d_0} a_i \chi(\Delta  (A_i)) + \sum_{j=1}^{d_1} b_j \chi(\Delta  (B_j)) = \varepsilon \chi(Z ).
\end{equation}
We define the function $S : V \rightarrow \R$ by
\begin{equation*}
S(x) =
\begin{cases}
\sum_{x \in A_i} a_i  & \text{ if } x \in X_0, \\
\sum_{x \in B_j} b_j  & \text{ if } x \in X_1.
\end{cases}
\end{equation*}

If $x \in X_0$ and $y \in X_1$ are arbitrary elements and $\{x,y\} \in Z$, then $\chi(\Delta  (A_i)) (\{x,y\}) = 1$
iff $x \in A_i$ and  $\chi(\Delta  (B_j)) (\{x,y\}) = 1$ iff $y \in B_j$. Therefore
for every $\{x,y\} \in Z$, the coordinate $\{x,y\}$
of~\cref{eq:dependent} gives

\begin{equation}
\label{eq:crossedge}
S(x) + S(y) = \varepsilon.
\end{equation}

From~\cref{eq:dependent} we can also deduce that for every $\{x, x'\} \in E(X_0)$ we have
\begin{equation}
\label{eq:zero0}
\sum_{i=1}^{d_0} a_i \chi(\Delta  (A_i)) (\{x, x'\} ) = 0,
\end{equation}
and for every $\{y, y'\} \in E(X_1)$ we have
\begin{equation}
\label{eq:zero1}
\sum_{j=1}^{d_1} b_j \chi(\Delta  (B_j)) (\{y,y'\}) = 0.
\end{equation}
Let $\{x_0,y_0\}$ be an arbitrary edge in $Z$, where $x_0 \in X_0$ and $y_0 \in X_1$. We set
$s_0 = S(x_0)$ and $s_1 = S(y_0)$. We know from~\cref{eq:crossedge} that
$$
s_0 + s_1 = \varepsilon .
$$
We claim that for every $\{x,y\} \in Z$, where $x \in X_0$ and $y \in X_1$, we have
$S(x) = s_0$ and $S(y) = s_1.$
For this consider an arbitrary breadth first search tree with root $x_0$.
Since the graph of the cut $Z$, the graph $G(Z) = (V', Z)$, is a connected bipartite graph,
every vertex in $V' \cap X_0$ will be at some even depth of the tree, and every vertex in $V' \cap X_1$
at some odd depth of the tree.
Going through all the vertices depth by depth starting with $x_0$ at
depth 0,~\cref{eq:crossedge} gives the claim.

We now distinguish two cases. In the first case at least one of $s_0$ and $s_1$ is non-zero,
say without loss of generality that $s_0 \neq 0$. For $i = 1, \ldots, d_0$, we define
\begin{equation*}
a'_i = a_i/s_0.
\end{equation*}
Then~\cref{eq:zero0} implies that in $G_0$, for every $\{x, x'\} \in E(X_0)$, we have
\begin{equation}
\label{eq:ok0}
\sum_{i=1}^{d_0} a'_i \chi(\Delta  (A_i)) (\{x, x'\} ) = 0.
\end{equation}
Also in $G_0$, if $x \in X_0$ then $\chi(\Delta  (A_i)) (\{x,v_1\}) = 1$  iff $x \in A_i$. Therefore
\begin{equation}
\label{eq:ok1}
\sum_{i=1}^{d_0} a'_i \chi(\Delta  (A_i)) (\{x, v_1\} ) = s_0/s_0 = 1.
\end{equation}
Therefore~\cref{eq:ok0,eq:ok1} imply that
\begin{equation}
\label{eq:dependent0}
\sum_{i=1}^{d_0} a'_i \chi(\Delta  (A_i))  = \chi(Z _0),
\end{equation}
which contradicts the linear independence of
$\{ \chi(\Delta  (A_1)), \ldots, \chi(\Delta  (A_{d_0})) , \chi(Z _0) \}$.

In the second case $s_0 = s_1 =0$, and thus for all $\{x,y\} \in Z$, with $x \in X_0$ and $y \in X_1$,
we have $S(x)=S(y)=0$. Therefore
in $G_0$, for every edge $\{x, v_1\}$,
\begin{equation}
\label{eq:aba0}
\sum_{i=1}^{d_0} a_i \chi(\Delta  (A_i)) (\{x, v_1\}) = {0},
\end{equation}
and similarly in $G_1$, for every edge $\{y, v_0\}$,
\begin{equation}
\label{eq:aba1}
 \sum_{j=1}^{d_1} b_j \chi(\Delta  (B_j)) (\{y, v_0\}) = {0}.
\end{equation}
Since $a_1, \ldots, a_{d_0}, b_1, \ldots , b_{d_1}$  are not all zero, either
$a_1, \ldots, a_{d_0}$  is not all zero or
$b_1, \ldots , b_{d_1}$ is not all zero. If $a_1, \ldots, a_{d_0}$  is not all zero then
from~\cref{eq:zero0,eq:aba0} it follows that
the family $\{ \chi(\Delta  (A_1)), \ldots, \chi(\Delta  (A_{d_0})) \}$ is dependent in $\laspan(\vec{\M}(G_0))$.
If $b_1, \ldots , b_{d_1}$ is not all zero then similarly
from~\cref{eq:zero1,eq:aba1} it follows that
the family $\{ \chi(\Delta  (B_1)), \ldots, \chi(\Delta  (B_{d_1})) \}$ is dependent in $\laspan(\vec{\M}(G_1))$. In either case, we reach a contradiction.
\end{proof}

\subsection{The upper bound}
We now can give our second proof of the upper bound on the cut dimension \cref{thm:cd_upper}.

\begin{proof}[Proof of \cref{thm:cd_upper}]
The proof is by induction.  For the base case $n=2$, the only graph to be considered consists
of a single edge and the cut dimension is $1 = 2n-3$.

Now let $n \ge 3$, and we assume the inductive hypothesis holds for all graphs on at most $n-1$ vertices.
We consider $3$ cases.

Case 1: The graph $G$ has only star mincuts, say at vertices $v_1, \ldots v_k$,
for some $1 \leq k \leq n$.  As there are only $k$ mincuts here we have $\cd(G) \le k \le n \le 2n-3$
for $n \ge 3$.

Case 2: There is a non-star mincut in $G$, and every non-star mincut is crossed by a non-star mincut.
Then by \cref{lem:crossing_cycle}, the graph $G$ is a cycle where the edges have all the same weight.
In this case by \cref{lem:cycledim}, we have $\cd(G) = \cd(C_n) = n \le 2n-3$ for $n \ge 3$.

Case 3 is where we use the induction hypothesis: Suppose that $G$ has a non-star crossless mincut $Z$
with shores $X_0$ and $X_1 = V \setminus X_0$.  Let $|X_0|=k$.
Then by \cref{lem:crosslessB} there are graphs $G_0,G_1$ such that $\cd(G) \le \cd(G_0) + \cd(G_1)-1$,
where $G_0$ is a graph on $k+1$ vertices, and $G_1$ is a graph on $n-k+1$ vertices. Therefore by the inductive hypothesis
\[
\cd(G) \le 2(k+1) - 3 + 2(n-k+1) -3 -1 = 2n-3 \enspace .
\]
\end{proof}

\subsection{The lower bound}
We now give our second proof of \cref{thm:cd_lower} that for every $n\ge 2$ there exist graphs $G$ with $\cd(G) = 2n-3$.
We need a slightly more detailed statement for the inductive hypothesis which is given in the following theorem.
\begin{theorem}
\label{thm:main_lower}
For every integer $n \ge 2$, there is a complete weighted graph $G = (V,w)$ on $n$ vertices with cut dimension $2n-3$
and minimum cut weight $1$, and where
for every $v \in V$, the star cut $\Delta(\{v\})$ is a minimum cut.
\end{theorem}

\begin{proof}
For $n=2$ the statement is satisfied by the graph consisting of  a single edge
of weight one which has cut dimension one and where the two star cuts are minimum cuts.  For $n=3$ we may
take the complete graph $G^{(3)} = (V^{(3)}, w^{(3)})$ with all weights $1/2$, which has cut dimension $3$.

Now assume that there exists a graph $G^{(n-1)} = (V^{(n-1)}, w^{(n-1)})$ on $n-1$ vertices satisfying the
inductive hypothesis. Let us consider a copy of $G^{(3)} = (V^{(3)}, w^{(3)})$ where
$V^{(3)} = \{t,u,v_0\}$ and
$ V^{(n-1)} \cap V^{(3)} = \emptyset$. We choose $v_1 \in V^{(n-1)}$ arbitrarily.
We claim that the $n$-vertex graph $G^n = (V^{(n)}, w^{(n)})$
defined as $\mer( \{ (G^{(n-1)}, v_1), (G^{(3)}, v_0) \} )$ satisfies the statement.
It follows from the definition of the merge operation that $G^{(n)}$ is a complete weighted graph and that its
star cuts are of weight one. In addition \cref{claim:sepmer} asserts that if
$Z$ is the cut in $G^n$ whose shores are
$V^{(n-1)}  \setminus \{v_1\}$ and $V^{(3)} \setminus \{v_0\}$ then $w(Z)=1$.

We now claim that the weight of a minimum cut of $G^{(n)} $ is one and that the mincut $Z$ is crossless.
Consider a non-star cut $\Delta(X)$. If both vertices $t,u$ are on the same shore then
the weight of $\Delta(X)$ is the same as the analogous cut in $G^{(n-1)} $ and therefore is at least one.
If $\Delta(X)$ crosses $Z$, then we suppose without loss of generality that $t \in X, u \in \bar{X}$.
We show that the weight of  $\Delta(X)$ is greater than one,
which then implies both claims.
The cut contains the edge $\{t,u\}$ which has weight $1/2$.
For every $y \in V^{(n-1)} \setminus \{v_1\},$ the cut
either contains the edge $\{t,y\}$ or the edge $\{u,y\}$, and these edges have the same weight.
Thus the total weight of such edges is half of the weight of $Z$, that is $1/2$.
In addition, the cut contains also at least one edge from $G^{(n-1)}$, therefore its total weight is greater than one.

Finally \cref{claim:sepmer} says that $\sep (G^{(n)}, Z) = \{G^{(n-1)}, G^{(3)} \}$. Since $Z$ is
a crossless non-star minimum cut that is also connected,
Lemma~\ref{lem:crosslessB} implies that $\cd(G^{(n)}) = \cd(G^{(n-1)}) + \cd(G^{(3)}) - 1$, which is $2n-3$
by the inductive hypothesis.
\end{proof}

\begin{figure}
\begin{centering}
\begin{tikzpicture}
\GraphInit[vstyle=Normal]
\begin{scope}
\Vertices[unit=2]{circle}{1,2,3,4,5,6,7,8}
\end{scope}
\SetUpEdge[color=red]
\Edge(1)(2)
\Edge(3)(4)
\Edge(5)(6)
\Edge(7)(8)
\SetUpEdge[color=black]
\Edge(1)(3)
\Edge(1)(7)
\Edge(2)(4)
\Edge(2)(6)
\Edge(3)(5)
\Edge(4)(8)
\Edge(5)(7)
\Edge(6)(8)
\end{tikzpicture}
\caption{Example graph $G$ showing the necessity of the connected condition in \cref{lem:crosslessA}.
Red edges have weight $2$ and black edges have weight $1$.  The minimum cut weight is 4 and the cuts
achieving this are all the star cuts and $\Delta(\{1,2\}), \Delta(\{3,4\}),\Delta(\{5,6\}),\Delta(\{7,8\}),\Delta(\{1,2,3,4\})$.}
\label{fig:8}
\end{centering}
\end{figure}
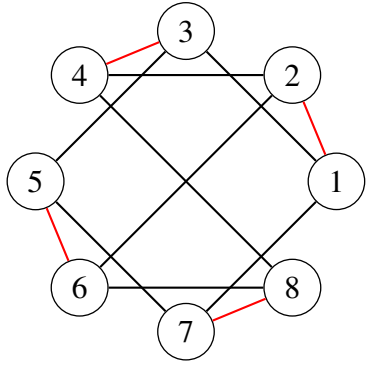

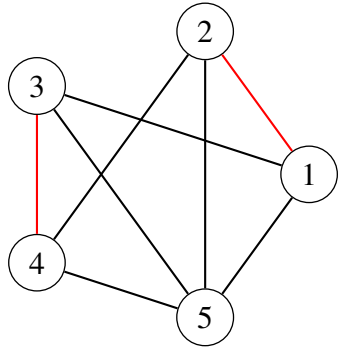
\begin{figure}
\begin{centering}
\begin{tikzpicture}
\GraphInit[vstyle=Normal]
\begin{scope}
\Vertices[unit=2]{circle}{1,2,3,4,5}
\end{scope}
\SetUpEdge[color=red]
\Edge(1)(2)
\Edge(3)(4)
\SetUpEdge[color=black]
\Edge(1)(3)
\Edge(1)(5)
\Edge(2)(4)
\Edge(2)(5)
\Edge(3)(5)
\Edge(4)(5)
\end{tikzpicture}
\caption{The graph $G_0$.  Red edges have weight $2$ and black edges have weight $1$.  The minimum cut weight is 4 and the cuts
achieving this are all the star cuts and $\Delta(\{1,2\}), \Delta(\{3,4\})$.  The cut dimension is 7.}
\label{fig:2}
\end{centering}
\end{figure}

\subsection{On the tightness of \cref{lem:crosslessB}}
One can wonder whether the connectedness of $Z$ is a necessary hypothesis in \cref{lem:crosslessB}.
In fact it is,
when $Z \in \M(G)$ is not connected then we can have $\cd(G) < \cd(G_0) + \cd(G_1)-1$.  An example is given in \cref{fig:8}.  The
mincuts in this graph are all the star cuts  and
\[
\Delta(\{1,2\}), \Delta(\{3,4\}),\Delta(\{5,6\}),\Delta(\{7,8\}),\Delta(\{1,2,3,4\}) \enspace.
\]
Thus no mincuts cross each other.  Also none of the non-star mincuts are connected.

Consider the case where $Z = \Delta(\{1,2,3,4\})$.  When we separate $G$ along this cut we see that $G_0 = G_1$ and they are
equal to the graph in \cref{fig:2}.  The mincuts in $G_0$ are all star cuts and $\Delta(\{1,2\}), \Delta(\{3,4\})$.  All non-star mincuts in
$G_0$ are connected so one can use \cref{lem:crosslessA} to compute that $\cd(G_0) = 7$, i.e.\ all these mincut vectors are linearly independent.
However, the cut dimension of $G$ is clearly at most 12 as it only has 12 edges.  Direct computation shows that in fact $\cd(G)=11$.

\section{$\ell_1$-approximate cut dimension}
\label{sec:l1-approx}
\label{sec:approx_construction}
In this section, we use the $\ell_1$-approximate cut dimension method to show \cref{thm:2n-2} that
for any $k \in \N$ and $n = 3k+1$, it holds that $D_\lin(\MINCUT_n) \ge 2n-2$.

Let $K_4$ be the complete graph on $4$ vertices with all edge weights equal to $1$.
The theorem will follow from showing that the $\ell_1$-approximate cut dimension of
the direct union of $k$ copies of $K_4$ has $\ell_1$-approximate cut dimension $6k$.
We start with the base case $k=1$ to build up the notation and intuition that will be needed
for the general case.  The following definition and fact will be useful.

\begin{definition}[Strictly diagonally dominant]
Let $A \in \R^{n \times n}$ be a matrix.  We say that the $\ith$ row of $A$ is \emph{strictly diagonally dominant}
if $|A(i,i)| > \sum_{j \ne i} |A(i,j)|$.  We say that $A$ is strictly diagonally dominant iff all of its rows are.
\end{definition}

It is well known that a strictly diagonally dominant matrix has full rank.
One way to prove this is via the following fact, which we will make use of in the proof of \cref{thm:2n-2}.
\begin{fact}
\label{fact:sdd}
Let $A \in \R^{n \times n}$ be a matrix whose $\ith$ row is strictly diagonally dominant.
If $A u = \zero$ for a vector $u \ne \zero$ then $|u_i| < \|u\|_\infty$.
\end{fact}

\begin{proof}
Suppose for a contradiction that for some $u \ne \zero$ it holds that $A u = \zero$ and $|u_i| = \|u\|_\infty$
where the $\ith$ row of $A$ is strictly diagonally dominant.
By normalizing and flipping the sign of $u$ if necessary we may assume $\|u\|_\infty =1$ and $A(i,i) u_i = |A(i,i)|$.  Thus
\begin{align*}
\sum_j A(i,j) u_j &= |A(i,i)| + \sum_{j \ne i} A(i,j) u_j \ge |A(i,i)| - \sum_{j \ne i} |A(i,j)| > 0 \enspace ,
\end{align*}
a contradiction.
\end{proof}

\subsection{$\ell_1$-approximate cut dimension of $K_4$}
\begin{figure}
\centering
\begin{tikzpicture}
\GraphInit[vstyle=Normal]
\tikzset{VertexStyle/.style = {draw, shape = circle, minimum size = 30pt}}
\Vertex[Math, x=0,y=0,L=v]{V}
\Vertex[Math,x=-2,y=2, L=a]{A}
\Vertex[Math,x=-4,y=0, L=b]{B}
\Vertex[Math,x=-2,y=-2, L=c]{C}
\Edge[label=3](V)(A)
\Edge[label=5, style={pos=.8}](V)(B)
\Edge[label=6](V)(C)
\Edge[label=1](A)(B)
\Edge[label=2,style={pos=.2}](A)(C)
\Edge[label=4](B)(C)
\end{tikzpicture}
\caption{The complete graph on $4$ vertices with all edge weights equal to $1$.  The labels on edges indicate the
ordering of edges used to represent cut vectors in the proof.}
\label{fig:K4}
\end{figure}
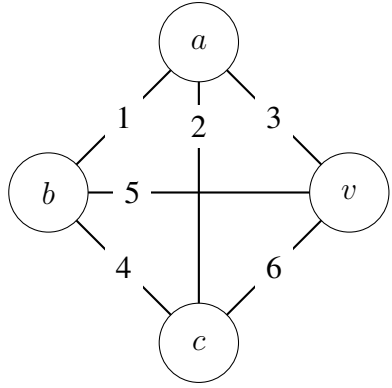

We label the vertices of $K_4$ by $a,b,c,v$, and
use the ordering of edges indicated in \cref{fig:K4}.
Let $X$ be the $7$-by-$6$ matrix whose rows correspond to the cut vectors of all the nontrivial cuts
\begin{equation}
\label{eq:baseX}
X =
\begin{bmatrix}
1 &  1 &  1 &  0 &  0 &  0 \\
1 &  0 &  0 &  1 &  1 &  0 \\
0 &  1 &  0 &  1 &  0 &  1 \\
0 &  0 &  1 &  0 &  1 &  1 \\
0 &  1 &  1 &  1 &  1 &  0 \\
1 &  0 &  1 &  1 &  0 &  1 \\
1 &  1 &  0 &  0 &  1 &  1 \\
\end{bmatrix}
\enspace .
\end{equation}
The cut vectors in $X$ are given in the order
\[
\Delta(\{a\}), \Delta(\{b\}), \Delta(\{c\}), \Delta(\{a,b,c\}), \Delta(\{a,b\}), \Delta(\{a,c\}), \Delta(\{b,c\}) \enspace.
\]
The first 4 rows correspond to star cuts which are minimum cuts of weight 3 in $K_4$.  The last three rows correspond to cuts which have weight 4 in $K_4$.
Thus to show a lower bound of $6$ on the number of linear queries needed to compute the minimum
cut of a 4 vertex graph, we need to show that the $w=(1,1,1,1,1,1), c=(0,0,0,0,1,1,1)$ one-sided $\ell_1$ approximate rank of $X$ is 6.

\begin{claim}
Let $w = \1 \in \R^6$, and $c =(0,0,0,0,1,1,1)$.  The $(w,c)$ one-sided $\ell_1$ approximate
rank of $X$ is $6$.
\end{claim}

\begin{proof}
The rank of $X$ at most $6$ as this is the number of columns, which takes care of the upper bound.

Now consider the lower bound.  To do this we need to lower bound the rank of the matrix
\[
Z=
X -
\begin{bmatrix}
\zero_{4,2} & \zero_{4,2} & \zero_{4,2} \\
A_1 & A_2 & A_3
\end{bmatrix}
\]
where each of $A_1, A_2,A_3 \ge 0$ are $3$-by-$2$ matrices and every row of $A_1 + A_2 + A_3$ sums to at most $1$.
As the first 4 rows of $X$ correspond to vectors of minimum cuts, no error is allowed on the first 4 rows.

The first 4 rows of $Z$ are equal to the first 4 rows of $X$, as there is no perturbation allowed on these rows.  By doing
elementary row operations on the first four rows, which do not change the rank, we can transform the first four rows of $Z$ into the reduced row echelon form
of $X(1:4, :)$.  Thus we arrive at the following matrix.
\[
\begin{bmatrix}
1 &  0 &  0 &  0 &  0 & -1 \\
0 &  1 &  0 &  0 & -1 &  0 \\
0 &  0 &  1 &  0 &  1 &  1 \\
0 &  0 &  0 &  1 &  1 &  1 \\
0 &  1 &  1 &  1 &  1 &  0 \\
1 &  0 &  1 &  1 &  0 &  1 \\
1 &  1 &  0 &  0 &  1 &  1 \\
\end{bmatrix}
-
\begin{bmatrix}
\zero_{4,2} & \zero_{4,2} & \zero_{4,2} \\
A_1 & A_2 & A_3
\end{bmatrix}
\enspace .
\]
Now we do column operations to zero out the entries in the first four rows and last two columns.  For a $m$-by-$2$ matrix $A$
we will use the notation $A^\circ$ to denote the matrix $A$ with the order of the columns swapped.
We arrive at
\[
\begin{bmatrix}
1 &  0 &  0 &  0 &  0 & 0 \\
0 &  1 &  0 &  0 & 0 &  0 \\
0 &  0 &  1 &  0 &  0 &  0 \\
0 &  0 &  0 &  1 &  0 &  0 \\
0 &  1 &  1 &  1 &  0 &  -2 \\
1 &  0 &  1 &  1 &  -2 &  0 \\
1 &  1 &  0 &  0 &  2 &  2 \\
\end{bmatrix}
-
\begin{bmatrix}
\zero_{4,2} & \zero_{4,2} & \zero_{4,2} \\
A_1 & A_2 & A_1^\circ -A_2 - A_2^\circ + A_3
\end{bmatrix}
\enspace .
\]
Finally, we can do row operations to zero out the first four columns in the last three rows.
\[
\begin{bmatrix}
1 &  0 &  0 &  0 &  0 & 0 \\
0 &  1 &  0 &  0 & 0 &  0 \\
0 &  0 &  1 &  0 &  0 &  0 \\
0 &  0 &  0 &  1 &  0 &  0 \\
0 &  0 &  0 &  0 &  0 &  -2 \\
0 &  0 &  0 &  0 &  -2 &  0 \\
0 &  0 &  0 &  0 &  2 &  2 \\
\end{bmatrix}
-
\begin{bmatrix}
\zero_{4,2} & \zero_{4,2} & \zero_{4,2} \\
\zero_{3,2} & \zero_{3,2} & A_1^\circ -A_2 - A_2^\circ + A_3
\end{bmatrix}
\enspace .
\]

The task has now reduced to showing the matrix
\[
Z'=
\begin{bmatrix}
2 & 0 \\
0 & 2 \\
-2 & -2
\end{bmatrix}
+ A_1 - A_2 - A_2^\circ + A_3^\circ
\]
has rank $2$ for any $A_1, A_2, A_3$ satisfying the constraints.  Let us simplify the matrix $A_1 - A_2 - A_2^\circ + A_3^\circ$.
First, let $A_1' = A_1 + A_3^\circ$.  Next, note that $D = A_2 + A_2^\circ$ has the property that $D(i,1) = D(i,2)$ for $i \in [3]$.
In the sequel we call this the \emph{partner property}.

As the row sum of $A_1' + A_2$ is at most $1$, unless $A_1'(1:2,1:2) = \zero_{2,2}$ and at least one row sum of $A_2(1:2,1:2)$ is equal to $1$ the
first two rows of $Z'$ will be strictly diagonally dominant.  If the first two rows of $Z'$ are strictly diagonally dominant then the
rank of $Z'$ must be $2$, thus we now handle the ``unless'' case.

First, suppose exactly one row sum of $A_2(1:2,1:2)$ is equal to $1$. Say without loss of generality it is the second one, thus
the first row of $Z'$ is strictly diagonally dominant.  Then for a sufficiently small $\varepsilon$ we can multiply the first column by $1-\varepsilon$
so that the first row remains strictly diagonally dominant and the second row becomes strictly diagonally dominant as well.  This
does not increase the rank and thus shows again that the rank of $Z'$ is $2$.

The remaining case is where both rows of $A_2(1:2,1:2)$ sum to one.  In this case by the partner property we have
\[
Z'(1:2,1:2) = \begin{bmatrix}
1&-1\\
-1&1
\end{bmatrix} .
\]
On the other hand, the last row of $Z'$ must have both entries $\le -1$.  Thus the determinant of the submatrix formed by the first row and the third is strictly negative and so $Z'$
has rank $2$.
\end{proof}

\subsection{Direct union of $K_4$ with itself}
\begin{figure}
\centering
\begin{minipage}{.55\textwidth}
\centering
\begin{tikzpicture}
\GraphInit[vstyle=Normal]
\tikzset{VertexStyle/.style = {draw, shape = circle, minimum size = 30pt}}
\Vertex[Math, x=0,y=0,L=v]{V}
\Vertex[Math,x=-2,y=2, L=a^{(1)}]{A}
\Vertex[Math,x=-4,y=0, L=b^{(1)}]{B}
\Vertex[Math,x=-2,y=-2, L=c^{(1)}]{C}
\Vertex[Math,x=2,y=2, L=a^{(2)}]{D}
\Vertex[Math,x=4,y=0, L=b^{(2)}]{E}
\Vertex[Math,x=2,y=-2, L=c^{(2)}]{F}
\Edge[label=3](V)(A)
\Edge[label=5, style={pos=.8}](V)(B)
\Edge[label=6](V)(C)
\Edge[label=9](V)(D)
\Edge[label=11, style={pos=.8}](V)(E)
\Edge[label=12](V)(F)
\Edge[label=1](A)(B)
\Edge[label=2,style={pos=.2}](A)(C)
\Edge[label=4](B)(C)
\Edge[label=7](D)(E)
\Edge[label=8,style={pos=.2}](D)(F)
\Edge[label=10](E)(F)
\end{tikzpicture}
\end{minipage}
\begin{minipage}{.4\textwidth}
\centering
\[
X^{(2)} =
\begin{bmatrix}
X & \zero_{7,6} \\
\zero_{7,6} & X
\end{bmatrix}
\]
\end{minipage}
\caption{Example of the direct union of two copies of $K_4$.  With the ordering of the edges given by the edge labels,
the matrix of cut vectors of the cuts $\Delta(\{a^{(i)}\}), \Delta(\{b^{(i)}\}), \Delta(\{c^{(i)}\}), \Delta(\{a^{(i)},b^{(i)},c^{(i)}\}), \Delta(\{a^{(i)},b^{(i)}\}), \Delta(\{a^{(i)},c^{(i)}\}), \Delta(\{b^{(i)},c^{(i)})\}$
for $i \in [2]$ becomes the matrix $X^{(2)}$ on the right.}
\label{fig:K42}
\end{figure}

Now we prove the general case.  The key to the proof is the following lemma.
\begin{lemma}
\label{lem:2k}
Let $k \in \N$ and $B$ be the $3k$-by-$2k$ matrix
\[
B = \begin{bmatrix} 2 \Id_{2k} \\ -2 \Id_k \otimes [1,1] \end{bmatrix} \enspace .
\]
For any matrices $3k$-by-$2k$ matrices $A_1, A_2$ satisfying the conditions
\begin{enumerate}
\item $A_1, A_2 \ge 0$
\item (partner property) For all $i \in [3k]$ and $j \in [k]$ it holds that $A_2(i,2j-1) = A_2(i,2j)$.
\item Every row of $A_1 + A_2/2$ sums to at most $1$
\end{enumerate}
it holds that $B +A_1 - A_2$ has rank $2k$.
\end{lemma}

\begin{proof}
The rank is at most $2k$ as that is the number of columns;  we focus on showing the columns are
linearly independent.

Let $Z = B + A_1 -A_2$.  We call the first $2k$ rows of $Z$ rows of type I, and the last $k$ rows
of type II.  If a type I row is not strictly diagonally dominant, we call it \emph{full}.  Notice that
a type I row $i$ is full if and only if the $\ith$ row of $A_1$ is zero and the $\ith$ row of $A_2$ sums to $2$.  In this
case, $Z(i,j) \le 0$ for every $j \ne i$ and it holds that $Z(i,i) = -\sum_{j \ne i} Z(i,j)$.
For $i \in [k]$ we call $2i-1$ and $2i$ \emph{partners}.

Suppose for contradiction there is a vector $\vecu \ne \zero$ such that $A\vecu = \zero$.  As $\vecu \ne \zero$ by
normalizing and multiplying by $-1$ as needed we may assume that $\|u\|_\infty = 1$ and $i$ is a coordinate with $\vecu(i)=1$.
By \cref{fact:sdd} the $\ith$ row of $Z$, which is a type I row, cannot be strictly diagonally dominant.  Thus the $\ith$ row
must be full.  Therefore for $Z(i,:) \vecu = 0$ to hold it must be the case that $\vecu(j) = 1$ for every $j$ where $A_2(i,j) > 0$.
Such a $j$ must exist as the $\ith$ row of $A_2$ sums to $2$.  So let $j$ be a coordinate with $A_2(i,j) > 0$ and let $j'$ be
the partner of $j$.  By the partner property we also have $A_2(i,j') > 0$ and therefore $\vecu(j) = \vecu(j') = 1$.

Now consider the type II row $\ell$ for which $B(\ell,j) = B(\ell,j') = -2$.
As $B(\ell,t) = 0$ for $t \not \in \{j,j'\}$ this means
\begin{align*}
Z(\ell,:) \vecu &= Z(\ell,j) + Z(\ell, j') + \sum_{t \not \in \{j,j'\}} Z(\ell,t) \vecu(t) \\
& \le B(\ell,j)+A_1(\ell,j) + B(\ell,j') + A_1(\ell,j') + \|\vecu\|_\infty \sum_{t \not \in \{j,j'\}} |Z(\ell,t)| \\
&\le -4 + \sum_{t} A_1(\ell,t) + A_2(\ell,t) \\
&\le -2 \enspace,
\end{align*}
and we have arrived at a contradiction.
 \end{proof}

With \cref{lem:2k} in hand we are now ready to prove \cref{thm:2n-2}.

\begin{proof}[Proof of \cref{thm:2n-2}]
Let $G^{(1)}, \ldots, G^{(k)}$ be $k$ copies of $K_4$ where the vertices in $G^{(i)}$ are labeled by $a^{(i)}, b^{(i)}, c^{(i)}, v^{(i)}$ for $i \in [k]$.
The graph $G$ is formed by taking the direct union of $G^{(1)}, \ldots, G^{(k)}$ at the vertices $v^{(1)}, \ldots, v^{(k)}$.  That is,
the vertices $v^{(1)}, \ldots, v^{(k)}$ are all identified by a common vertex denoted $v$.  See \cref{fig:K42} for an illustration of the graph
for $k=2$.

The cuts of $G$ we focus on are the $7k$ cuts given by
\[
\Delta(\{a^{(i)}\}), \Delta(\{b^{(i)}\}), \Delta(\{c^{(i)}\}), \Delta(\{a^{(i)},b^{(i)},c^{(i)}\}), \Delta(\{a^{(i)},b^{(i)}\}), \Delta(\{a^{(i)},c^{(i)}\}), \Delta(\{b^{(i)},c^{(i)}\} \enspace,
\]
for $i \in [k]$.  For any $i \in [k]$ the cuts $\Delta(\{a^{(i)}\}), \Delta(\{b^{(i)}\}), \Delta(\{c^{(i)}\}), \Delta(\{a^{(i)},b^{(i)},c^{(i)}\})$ achieve the minimum cut weight of $G$, which is 3,
and the cuts $\Delta(\{a^{(i)},b^{(i)}\}), \Delta(\{a^{(i)},c^{(i)}\}), \Delta(\{b^{(i)},c^{(i)}\}$ have weight 4.

With an ordering of the edges as exemplified in \cref{fig:K42}, the matrix of cut vectors of these cuts is $X^{(k)} = \Id_k \otimes X$,
where $X$ is the matrix from \cref{eq:baseX}.  In every nonzero block of $X^{(k)}$ the first four rows are minimum cuts with weight 3 and the last 3 rows are cuts with weight 4.
Let $c' = (0,0,0,0,1,1,1)$.  The theorem will follow from \cref{thm:approx_cut} by showing that the $w = \mathbf{1}_{6k}, c = \mathbf{1}_k \otimes c'$ one-sided $\ell_1$ approximate
rank of $X^{(k)}$ is $6k$.

To do this, we must show that $X^{(k)} - A$ has rank $6k$ for any matrix $A \ge 0$ which is all zero on any row of $I_k \otimes X$ corresponding
to a minimum cut, and where the row sum of $A$ is at most $1$ on any row of $I_k \otimes X$ corresponding to a cut of weight $4$.
In order to make reference to the base case, it will be useful to partition the columns into $k$ blocks of $6$ columns, where the $\ith$ block
is further partitioned into blocks of size $2$ represented by the $7k$-by-$2$ matrices $A_1^{(i)}, A_2^{(i)},A_3^{(i)}$.  In other words, we view
$A$ as follows
\[
A =
\begin{bmatrix}
A_1^{(1)} & A_2^{(1)} & A_3^{(1)} & \cdots & A_1^{(k)} & A_2^{(k)} & A_3^{(k)}
\end{bmatrix}
\enspace
\]
where each $A_j^{(i)}$ for $j \in [3], i \in [k]$ is a $7k$-by-2 matrix.

As in the base case, we begin by doing Gauss-Jordan elimination on the rows corresponding to mincuts of each $X$ block in $X^{(k)}$.  These operations
only touch rows corresponding to mincuts where $A$ is zero, thus they do not change $A$.  After these operations we arrive at the matrix $\Id_k \otimes X' - A$
where
\[
X' =
\begin{bmatrix}
1 &  0 &  0 &  0 &  0 & -1 \\
0 &  1 &  0 &  0 & -1 &  0 \\
0 &  0 &  1 &  0 &  1 &  1 \\
0 &  0 &  0 &  1 &  1 &  1 \\
0 &  1 &  1 &  1 &  1 &  0 \\
1 &  0 &  1 &  1 &  0 &  1 \\
1 &  1 &  0 &  0 &  1 &  1 \\
\end{bmatrix}
\]
Next, as in the base case, we do column operations to zero out the last two columns in the first four rows of each block of $X'$.  This
gives us the matrix $\Id_k \otimes X'' - A'$ where
\[
X'' =
\begin{bmatrix}
1 &  0 &  0 &  0 &  0 & 0 \\
0 &  1 &  0 &  0 & 0 &  0 \\
0 &  0 &  1 &  0 &  0 &  0 \\
0 &  0 &  0 &  1 &  0 &  0 \\
0 &  1 &  1 &  1 &  0 &  -2 \\
1 &  0 &  1 &  1 &  -2 &  0 \\
1 &  1 &  0 &  0 &  2 &  2 \\
\end{bmatrix}
\]
and the $\ith$ block of $A'$ looks like
\[
\begin{tabular}{c|c|c}
$[A_1^{(i)}$ & $A_2^{(i)}$ & $A_1^{(i)\circ} -A_2^{(i)} - A_2^{(i)\circ} + A_3^{(i)}] .$
\end{tabular}
\enspace.
\]
Here $A_1^{(i)\circ}$ denotes the matrix $A_1^{(i)}$ with the order of the columns swapped.
Finally, we use $X''(1:4,1:4)$ to zero out all other entries of $\Id_k \otimes X'' - A'$ in the first 4 columns of each block.
This brings us to the matrix $\Id_k \otimes X''' - A''$ where
\[
X''' =
\begin{bmatrix}
1 &  0 &  0 &  0 &  0 & 0 \\
0 &  1 &  0 &  0 & 0 &  0 \\
0 &  0 &  1 &  0 &  0 &  0 \\
0 &  0 &  0 &  1 &  0 &  0 \\
0 &  0 &  0 &  0 &  0 &  -2 \\
0 &  0 &  0 &  0 &  -2 &  0 \\
0 &  0 &  0 &  0 &  2 &  2 \\
\end{bmatrix}
\]
and the $\ith$ block of $A''$ is
\[
\begin{tabular}{c|c|c}
$[\zero_{7k,2}$ & $\zero_{7k,2}$ & $A_1^{(i)\circ} -A_2^{(i)} - A_2^{(i)\circ} + A_3^{(i)}]$
\end{tabular}
\enspace.
\]
Again, each of $A_1^{(i)}, A_2^{(i)}, A_3^{(i)}$ is zero on rows corresponding to minimum cuts.  Thus by multiplying the last two columns of
each block by $-1$ and permuting rows and columns we can transform $\Id_k \otimes X''' - A''$ into the form
\[
\begin{bmatrix}
\Id_{4k} & \zero_{4k,2k} \\
\zero_{3k,4k} & B +A_1 - A_2
\end{bmatrix}
\]
where $B,A_1, A_2$ satisfy the conditions of \cref{lem:2k}.  Thus a rank lower bound of $6k$ follows from the lower
bound of $2k$ on the rank of $B+A_1-A_2$ given in \cref{lem:2k}.
\end{proof}

\section{The dimension of approximate mincuts}\label{sec:alphacuts}
Let $G$ be a weighted graph and $\lambda$ the weight of a minimum cut in $G$.
For $\alpha \ge 1$ define an $\alpha$-near-mincut of $G$ to be a cut $S$ whose weight is at most $\alpha \lambda$.
Let $\M_\alpha(G)$ be the set of all $\alpha$-near-mincuts of $G$ and $\vec{\M}_\alpha(G) = \{\chi(S): S \in \M_\alpha(G)\}$.
In this section, we look at $\cd_\alpha(G) = \dim(\laspan(\vec{\M}_\alpha(G)))$.

The first observation is that if $\alpha=2$ then the unweighted complete graph $K_n$ satisfies $\cd_\alpha(K_n) = \binom{n}{2}$.
For \emph{simple} graphs we can show $\alpha=2$ is a sharp threshold.
\begin{theorem}
\label{thm:alpha}
Let $1\leq \alpha < 2$ be a constant and $G$ be a simple $n$-vertex graph.  Then $\cd_\alpha(G) = O(n)$.
\end{theorem}

The key to this theorem is the following lemma of Rubinstein, Schramm, and Weinberg \cite{RSW18}.\kern-3mm
\begin{lemma}[Lemma 2.6 \cite{RSW18}]
\label{lem:RSW}
Let $G$ be a simple graph with minimum degree $d_{\min}$ and minimum cut value $\lambda$.  For constant
$0\leq\epsilon < 1$ let $\Tcal$ be the set of non-star cuts of $G$ whose weight is at most $\lambda + \epsilon d_{\min}$.
Then $| \cup_{T \in \Tcal} T | = O(n)$.
\end{lemma}

\begin{proof}[Proof of \cref{thm:alpha}]
Let $G$ be a simple graph.  To prove the theorem we create a set of $O(n)$ vectors that span $\vec{\M}_\alpha(G)$.
Let $\M_\alpha(G) = \Tcal \sqcup \Scal$, where $\Tcal$ is the set of non-star cuts of $\M_\alpha(G)$ and $\Scal$ is the set of star cuts of $\M_\alpha(G)$.
Let $E' = \cup_{T \in \Tcal} T$ be the set of edges involved in the cuts in $\Tcal$.  Let $\vec{\Lcal} = \{e_i : i \in E'\}$.
Note that from the definition of $d_{\min}$, there is a star cut with cut value $d_{\min}$, which implies that $\lambda \leq d_{\min}$. As a result, every $\alpha$-{near-}mincut has cut value at most
$\alpha \lambda \leq \lambda+ (\alpha-1) d_{\min}$, and hence by \cref{lem:RSW} we have $|\vec{\Lcal}| = O(n)$.
Also $\laspan(\vec{\Tcal}) \subseteq \laspan(\vec{\Lcal})$.  Thus
$\laspan(\vec{\M}_\alpha(G)) \subseteq \laspan(\vec{\Lcal} \cup \vec{\Scal})$.  As $|\Scal| \le n$ this is a spanning set of size $O(n)$.
\end{proof}

In a previous version of this work we conjectured that for an $n$-vertex \emph{weighted} graph $G$ it holds that $\cd_\alpha(G) = O(n)$ for any $\alpha < 4/3$.  This turns out to be
false, however.  The reason is that, on the one hand, in a graph $G=(V,w)$ the characteristic vector of a cut $\chi(S)$ depends only on the set of edges, but not the weight of these edges.  On the
other hand, $w(S)$ does of course depend on the weight of the edges. We can utilize this difference to construct an example as follows.  Let us start with a cycle $C_n$ with all edge weights being 1.  While $C_n$ has $\binom{n}{2}$ mincuts 
with weight 2, these mincuts live in an
$n$-dimensional space as $C_n$ only has $n$ edges.  We can then turn $C_n$ into a complete weighted graph $G$ by adding a tiny weight $\varepsilon = 2(\alpha-1)/\binom{n}{2}$
edge to all pairs of vertices that are not adjacent in the cycle.  As adding edges cannot decrease the minimum cut weight, the weight of a minimum cut in $G$ is at least $2$.  Further, if $X$ is the 
shore of a minimum cut in $C_n$ then in the graph $G$ we have $w(\Delta(X)) \le 2 + \binom{n}{2} \varepsilon = 2\alpha$, as the weight is at most its weight in $C_n$ plus the weight of all added edges.  
Thus $\Delta(X)$ is an $\alpha$-near-mincut in $G$.
Further, the characteristic vectors $\chi(\Delta(X)) \in \{0,1\}^{\binom{n}{2}}$ of these cuts in $G$ now live in an $\binom{n}{2}$-dimensional space and
become linearly independent. This example demonstrates that a reasonable extension of the cut dimension to near-mincuts should take into account the magnitude of the edge weights,
as the $\ell_1$-approximate cut dimension does.

We now give the formal proof that the graph $G$ mentioned above has the correct properties.
\begin{lemma}
\label{lem:dimCshore}
Let $n \in \N$.  Let $C_n$ be the cycle on $n$ vertices and $\Gcal$ the beach of $\M(C_n)$.
Let $K_n$ be the complete graph on $n$ vertices. Let $\Tcal = \{\Delta(X): X \in \Gcal\}$, where
here $\Delta(X) \in \{0,1\}^{\binom{n}{2}}$ is the cut in $K_n$ with shore $X$.
Then $\dim(\laspan(\vec{\Tcal})) = \binom{n}{2}$.
\end{lemma}

\begin{proof}
For this proof we assume the vertices are labeled by $0,\ldots, n-1$ and use addition modulo $n$.
We will show that all of the standard basis vectors $e_{\{i,j\}}$ are in $\laspan(\vec{\Tcal})$.  For concreteness,
we show how to construct the vectors $e_{\{0,j\}}$; by symmetry the same argument can then be used for any $e_{\{i,j\}}$.

We will actually construct the vectors $E_j = \sum_{k=1}^j e_{\{0,k\}}$.  This suffices as $e_{\{0,j\}} = E_j - E_{j-1}$.
First note that $e_{\{0,1\}} = \frac{1}{2}(\chi(\Delta(\{0\}) + \chi(\Delta(\{1\})) - \chi(\Delta(\{0,1\})))$,
and thus is in $\laspan(\vec{\Tcal})$ as all the vectors on the right hand side are in $\vec{\Tcal}$.

Now let $j >1$ and $X = \{1, \ldots, j\}, X' = X \cup \{0\}$. Then
\[
\chi(\Delta(X))(e) - \chi(\Delta(X'))(e) =
\begin{cases}
1 & \mbox{ if } e = \{0, k\}, k \in X \\
-1 & \mbox{ if } e = \{0,k\}, k \in \bar{X'} \\
0 & \mbox{ otherwise}
\end{cases} \enspace .
\]
Thus $E_j = \frac{1}{2}(\Delta(\{0\} + \chi(\Delta(X))- \chi(\Delta(X')))$.
\end{proof}

\begin{theorem}
Let $n \in \N$.  For any $\alpha > 1$ there exists a graph $G = (\{0,\ldots, n-1\},w)$ such that $\cd_\alpha(G) = \binom{n}{2}$.
\end{theorem}

\begin{proof}
We again use addition modulo $n$ on the labels of the vertices.
Let $\varepsilon = 2(\alpha-1)/\binom{n}{2}$.  Define $w(\{i,i+1\}) = 1$ for $i \in \{0,\ldots,n-1\}$ and for any
other $i,j$ let $w(\{i,j\}) = \varepsilon$.  Let $G = (\{0,\ldots, n-1\},w)$.  Thus $G$ is the graph of the cycle $C_n$ with edges of weight
$\varepsilon$ added between all pairs of vertices that are not adjacent in the cycle.  The weight of a minimum cut of $G$ is at least that of 
$C_n$, which is $2$, as adding edges cannot decrease the weight of a cut.  Further, if $X$ is the 
shore of a minimum cut in $C_n$ then in the graph $G$ we have $w(\Delta(X)) \le 2 + \binom{n}{2} \varepsilon = 2\alpha$, as the weight is at most its weight in $C_n$ plus the weight of all added edges.  
Thus $\Delta(X)$ is an $\alpha$-near-mincut in $G$ and $\cd_\alpha(G)$ is at least $\binom{n}{2}$ by \cref{lem:dimCshore}.  It also clearly cannot be larger 
than $\binom{n}{2}$ and so the theorem is proved.
\end{proof}


\section*{Acknowledgments}
Troy Lee is supported in part by the Australian Research Council Grant No: DP200100950.
Research at CQT is funded by the National Research Foundation, the Prime Minister's Office, and the Ministry of Education, Singapore under the Research Centres of Excellence programme's research grant R-710-000-012-135.
In addition, this work has been supported in part by the QuantERA ERA-NET Cofund project QuantAlgo and the ANR project ANR-18-CE47-0010 QUDATA.
Tongyang Li is supported by the ARO contract W911NF-17-1-0433, NSF grant PHY-1818914, and an NSF QISE-NET Triplet Award (grant DMR-1747426).


\begin{thebibliography}{GMW20b}

\bibitem[ACK20]{ACK20}
Sepehr Assadi, Deeparnab Chakrabarty, and Sanjeev Khanna.
\newblock Graph connectivity and single element recovery via linear and {OR}
  queries.
\newblock {\em CoRR}, abs/2007.06098, 2020.

\bibitem[BFS86]{BFS86}
L{\'{a}}szl{\'{o}} Babai, Peter Frankl, and Janos Simon.
\newblock Complexity classes in communication complexity theory (preliminary
  version).
\newblock In {\em 27th Annual Symposium on Foundations of Computer Science,
  Toronto, Canada, 27-29 October 1986}, pages 337--347, 1986.

\bibitem[BG08]{BG08}
Andr{\'{a}}s~A. Bencz{\'{u}}r and Michel~X. Goemans.
\newblock Deformable polygon representations and near-mincuts.
\newblock In Martin Gr{\o}tschel and Gyula O.~H. Katona, editors, {\em Building
  Bridges: Between Mathematics and Computer Science}, volume~19 of {\em Bolyai
  Society Mathematical Studies}, pages 103--135. Springer, 2008.

\bibitem[Bix75]{bixby75}
R.~E. Bixby.
\newblock The minimum number of edges and vertices in a graph with edge
  connectivity n and m n-bonds.
\newblock {\em Netw.}, 5(3):253–298, July 1975.

\bibitem[CR04]{CR04}
L.~Sunil Chandran and L.~Shankar Ram.
\newblock On the number of minimum cuts in a graph.
\newblock {\em {SIAM} J. Discret. Math.}, 18(1):177--194, 2004.

\bibitem[DKL76]{DKL76}
Efim~A. Dinitz, Alexander~V. Karzanov, and Michael~V. Lomonosov.
\newblock On the structure of the system of minimum edge cuts of a graph.
\newblock {\em Studies in discrete optimization}, 1976.

\bibitem[FF09]{FF09}
Tam{\'{a}}s Fleiner and Andr{\'{a}}s Frank.
\newblock A quick proof for the cactus representation of mincuts.
\newblock {\em EGRES Quick Proof}, 2009-03, 2009.

\bibitem[GMW20a]{GMW20}
Pawel Gawrychowski, Shay Mozes, and Oren Weimann.
\newblock Minimum cut in ${O}(m \log^2 n)$ time.
\newblock In Artur Czumaj, Anuj Dawar, and Emanuela Merelli, editors, {\em 47th
  International Colloquium on Automata, Languages, and Programming, {ICALP}
  2020, July 8-11, 2020, Saarbr{\"{u}}cken, Germany (Virtual Conference)},
  volume 168 of {\em LIPIcs}, pages 57:1--57:15. Schloss Dagstuhl -
  Leibniz-Zentrum f{\"{u}}r Informatik, 2020.

\bibitem[GMW20b]{GMW20b}
Pawel Gawrychowski, Shay Mozes, and Oren Weimann.
\newblock A note on a recent algorithm for minimum cut.
\newblock {\em CoRR}, abs/2008.02060, 2020.

\bibitem[GNT20]{GNT20}
Mohsen Ghaffari, Krzysztof Nowicki, and Mikkel Thorup.
\newblock Faster algorithms for edge connectivity via random 2-out
  contractions.
\newblock In Shuchi Chawla, editor, {\em Proceedings of the 2020 {ACM-SIAM}
  Symposium on Discrete Algorithms, {SODA} 2020, Salt Lake City, UT, USA,
  January 5-8, 2020}, pages 1260--1279. {SIAM}, 2020.

\bibitem[Goe06]{Goemans06}
Michel~X. Goemans.
\newblock Minimum bounded degree spanning trees.
\newblock In {\em 47th Annual {IEEE} Symposium on Foundations of Computer
  Science {(FOCS} 2006), 21-24 October 2006, Berkeley, California, USA,
  Proceedings}, pages 273--282. {IEEE} Computer Society, 2006.

\bibitem[GPRW20]{GPRW20}
Andrei Graur, Tristan Pollner, Vidhya Ramaswamy, and S.~Matthew Weinberg.
\newblock New query lower bounds for submodular function minimization.
\newblock In Thomas Vidick, editor, {\em 11th Innovations in Theoretical
  Computer Science Conference, {ITCS} 2020, January 12-14, 2020, Seattle,
  Washington, {USA}}, volume 151 of {\em LIPIcs}, pages 64:1--64:16. Schloss
  Dagstuhl - Leibniz-Zentrum f{\"{u}}r Informatik, 2020.

\bibitem[GR95]{GR95}
Michel~X. Goemans and V.~S. Ramakrishnan.
\newblock Minimizing submodular functions over families of sets.
\newblock {\em Comb.}, 15(4):499--513, 1995.

\bibitem[Har08]{Harvey08}
Nicholas J.~A. Harvey.
\newblock {\em Matchings, matroids and submodular functions}.
\newblock PhD thesis, Massachusetts Institute of Technology, Cambridge, MA,
  {USA}, 2008.

\bibitem[HMT88]{HMT88}
Andr{\'{a}}s Hajnal, Wolfgang Maass, and Gy{\"{o}}rgy Tur{\'{a}}n.
\newblock On the communication complexity of graph properties.
\newblock In {\em Proceedings of the 20th Annual {ACM} Symposium on Theory of
  Computing, May 2-4, 1988, Chicago, Illinois, {USA}}, pages 186--191, 1988.

\bibitem[HRW20]{HRW20}
Monika Henzinger, Satish Rao, and Di~Wang.
\newblock Local flow partitioning for faster edge connectivity.
\newblock {\em {SIAM} J. Comput.}, 49(1):1--36, 2020.

\bibitem[HW96]{HW96}
Monika~Rauch Henzinger and David~P. Williamson.
\newblock On the number of small cuts in a graph.
\newblock {\em Inf. Process. Lett.}, 59(1):41--44, 1996.

\bibitem[Jai01]{Jain01}
Kamal Jain.
\newblock A factor 2 approximation algorithm for the generalized {S}teiner
  network problem.
\newblock {\em Comb.}, 21(1):39--60, 2001.

\bibitem[Kar93]{Karger93}
David~R. Karger.
\newblock Global min-cuts in {RNC}, and other ramifications of a simple min-cut
  algorithm.
\newblock In Vijaya Ramachandran, editor, {\em Proceedings of the Fourth Annual
  {ACM/SIGACT-SIAM} Symposium on Discrete Algorithms, 25-27 January 1993,
  Austin, Texas, {USA}}, pages 21--30. {ACM/SIAM}, 1993.

\bibitem[Kar00]{Karger00}
David~R. Karger.
\newblock Minimum cuts in near-linear time.
\newblock {\em J. {ACM}}, 47(1):46--76, 2000.

\bibitem[KT19]{KT19}
Ken{-}ichi Kawarabayashi and Mikkel Thorup.
\newblock Deterministic edge connectivity in near-linear time.
\newblock {\em J. {ACM}}, 66(1):4:1--4:50, 2019.

\bibitem[KV18]{KV18}
Bernhard Korte and Jens Vygen.
\newblock {\em Combinatorial Optimization: Theory and Algorithms}.
\newblock Springer, 2018.

\bibitem[Lov93]{Lov93}
L{\'{a}}szl{\'{o}} Lov{\'{a}}sz.
\newblock {\em Combinatorial problems and exercises {(2.} ed.)}.
\newblock North-Holland, 1993.

\bibitem[MN20]{MN20}
Sagnik Mukhopadhyay and Danupon Nanongkai.
\newblock Weighted min-cut: sequential, cut-query, and streaming algorithms.
\newblock In {\em Proceedings of the 52nd Annual ACM SIGACT Symposium on Theory
  of Computing, STOC 2020, Chicago, IL, USA, June 22-26, 2020}, pages 496--509,
  2020.

\bibitem[NNI97]{NNI97}
Hiroshi Nagamochi, Kazuhiro Nishimura, and Toshihide Ibaraki.
\newblock Computing all small cuts in an undirected network.
\newblock {\em {SIAM} J. Discret. Math.}, 10(3):469--481, 1997.

\bibitem[RSW18]{RSW18}
Aviad Rubinstein, Tselil Schramm, and S.~Matthew Weinberg.
\newblock Computing exact minimum cuts without knowing the graph.
\newblock In {\em 9th Innovations in Theoretical Computer Science Conference,
  {ITCS} 2018, January 11-14, 2018, Cambridge, MA, {USA}}, pages 39:1--39:16,
  2018.

\end{thebibliography}



\appendix
\section{Jain's spanning lemma}
\label{sec:appendix}
In this appendix we prove \cref{thm:span}.  The proof uses the following key property of mincuts which goes back at least
to work of Dinitz, Karzanov, and Lomonosov \cite{DKL76}.

\begin{proposition}[\cite{DKL76} ``Lemma on a quadrangle'']
\label{prop:bixby}
Let $G = (V,w)$ be a graph.
For any crossing mincuts $\Delta(X), \Delta(Y)$ of $G$
it holds that
\[
\chi(\Delta(X)) + \chi(\Delta(Y)) = \chi(\Delta(X \cap Y)) + \chi(\Delta(X \cup Y)) \enspace .
\]
\end{proposition}

\begin{proof}
If $\Delta(X), \Delta(Y)$ cross then
$\Delta(X \cap Y), \Delta(X \cup Y)$ are mincuts of $G$ by \cref{claim:mincuts}.
Further, by counting the number of times an edge appears on each side it can be seen (eg.\ Ex. 6.48 in \cite{Lov93}) that
\begin{equation}
\label{eq:cut_vecs}
\chi(\Delta(X)) + \chi(\Delta(Y)) = \chi(\Delta(X \cap Y)) + \chi(\Delta(X \cup Y)) + 2\chi(E(X-Y, Y-X)) \enspace.
\end{equation}

Let the minimum cut value of $G$ be $\lambda$.  Let $m$ be the number of edges in $G$ and
$\vec{w} \in \R^m$ be the positive vector resulting from restricting $w$ to the edges of $G$.
The inner product of $\vec{w}$ with the left hand side of \cref{eq:cut_vecs}
is $2\lambda$, and with the righthand side is $2\lambda  + 2 \braket{\vec{w}}{\chi(E(X-Y, Y-X))}$.  Thus $\braket{\vec{w}}{\chi(E(X-Y, Y-X))} = 0$,
which implies $\chi(E(X-Y, Y-X)) = \zero$ since
$\vec{w}$ is positive and $\chi(E(X-Y, Y-X))$ is nonnegative.
\end{proof}

Jain's proof uses the technique of \emph{combinatorial uncrossing}.
Recall the definition of $\cross_{\Gcal}(X)$ from \cref{def:overlap}.   A key to the proof is the following simple lemma about
$\cross_{\Gcal}(X)$.
\begin{lemma}[\cite{Jain01}]
\label{lem:intersect}
Let $\Fcal \subseteq 2^{V}$ be a set family closed under overlaps and $\Gcal \subseteq \Fcal$ be a maximal laminar subset of $\Fcal$.
Then for any $X \in \Fcal - \Gcal$ and $Y \in \cross_{\Gcal}(X)$
\begin{align}
\label{eq:con_cap}
\cross_{\Gcal}(X \cap Y) & \subset \cross_{\Gcal}(X) \\
\label{eq:con_cup}
\cross_{\Gcal}(X \cup Y) &\subset \cross_{\Gcal}(X) \enspace .
\end{align}
\end{lemma}

\begin{proof}
In the following we always refer to $\cross(X)$ with respect to $\Gcal$ and drop the subscript.
We first show \cref{eq:con_cap}.  First note that $Y \in \cross(X) - \cross(X \cap Y)$.  Thus to show \cref{eq:con_cap} it suffices to show $\cross(X \cap Y) \subseteq \cross(X)$.
Let $W \in \cross(X \cap Y)$.  We want to show that $W \in \cross(X)$, i.e.\ that it cannot be the case that $W \subseteq X, X \subseteq W$, or
$X \cap W = \emptyset$.  We know that the last one cannot hold as $W \cap (X \cap Y) \ne \emptyset$ as $W \in \cross(X \cap Y)$.

Also as $W,Y \in \Gcal$ they do not overlap and thus either $Y \subseteq W, W \subseteq Y$, or $Y \cap W = \emptyset$.  Again the last one
cannot hold as $W \cap (X \cap Y) \ne \emptyset$.  The following table shows that assuming $W \not \in \cross(X)$ leads to a contradiction
in all $4$ remaining cases.
\begin{center}
\begin{tabular}{c|c|c|}
                                       &      $Y \subseteq W$       & $W \subseteq Y$                              \\
\hline
$W \subseteq X$
& \begin{tabular}{@{}c@{}} $Y \subseteq X$ \\ $Y \not \in \cross(X)$ \end{tabular}
& \begin{tabular}{@{}c@{}} $W \subseteq X \cap Y$ \\ $W \not \in \cross(X \cap Y)$ \end{tabular}   \\
\hline
$X \subseteq W$
&  \begin{tabular}{@{}c@{}} $X \cap Y \subseteq W$ \\ $W \not \in \cross(X \cap Y)$ \end{tabular}
& \begin{tabular}{@{}c@{}} $X \subseteq Y$ \\ $Y \not \in \cross(X)$ \end{tabular}  \\
\hline
\end{tabular}
\end{center}

We now show \cref{eq:con_cup}, which follows similarly.  Again $Y \in \cross(X) - \cross(X \cup Y)$ thus it suffices to show
$\cross(X \cup Y) \subseteq \cross(X)$.  Let $W \in \cross(X \cup Y)$.  We want to show that $W \in \cross(X)$, i.e.\ that
is not the case that either $W \cap X = \emptyset, X \subseteq W$, or $W \subseteq X$.  We cannot have $W \subseteq X$
because this means $W \subseteq X \cup Y$ which contradicts $W \in \cross(X \cup Y)$.  As $W,Y \in \Gcal$ they do
not overlap, so we also know either $Y \subseteq W, W \cap Y = \emptyset$, or $W \subseteq Y$.  The last one again
cannot hold as it implies $W \subseteq X \cup Y$.  The following table shows that assuming $W \not \in \cross(X)$ leads
to a contradiction in the remaining $4$ cases.
\begin{center}
\begin{tabular}{c|c|c|}
                                       &      $Y \subseteq W$            & $W \cap Y = \emptyset$ \\
\hline
$X \subseteq W$
& \begin{tabular}{@{}c@{}} $X \cup Y \subseteq W$ \\ $W \not \in \cross(X \cup Y)$ \end{tabular}
& \begin{tabular}{@{}c@{}} $X \cap Y = \emptyset$ \\ $Y \not \in \cross(X)$ \end{tabular} \\
\hline
$X \cap W = \emptyset$
&  \begin{tabular}{@{}c@{}} $Y \cap X = \emptyset$ \\ $Y \not \in \cross(X)$ \end{tabular}
& \begin{tabular}{@{}c@{}} $W \cap (X \cup Y) = \emptyset$ \\ $W \not \in \cross(X \cup Y)$ \end{tabular} \\
\hline
\end{tabular}
\end{center}
\end{proof}

We are now ready to show the key lemma of Jain.
\jainspan*

\begin{proof}
It is clear that $\laspan(\vec{\Lcal}) \subseteq \laspan(\vec{\M}(G))$ so we focus on the other direction.

Let $\Fcal$ be the beach of $\M(G)$.  By \cref{claim:mincuts} $\Fcal$ is closed under overlaps.  Let $\Gcal \subseteq \Fcal$ be the beach of $\Lcal$.  As $\Lcal$ is a
maximal cross-free subset of $\M(G)$ it follows that $\Gcal$ is a maximal laminar subset of $\Fcal$.
Thus $|\cross_{\Gcal}(X)| \ge 1$ for all $X \in \Fcal - \Gcal$.  In the following we will always refer to $\cross(X)$ with respect to $\Gcal$ and drop the subscript.

Suppose for a contradiction that $\laspan(\vec{\Lcal}) $ is a strict subset of
$\laspan(\vec{\Mcal}(G))$.  Let
\[
X = \argmin_{Z \in \Fcal - \Gcal} \{ |\cross(Z) |:  \chi(\Delta(Z)) \not \in  \laspan(\vec{\Lcal}) \} \enspace .
\]
As $\cross(X) \ge 1$, let $Y \in \cross(X)$.  By \cref{lem:intersect}
\begin{align}
\label{eq:card_cap}
|\cross(X \cap Y)| &< |\cross(X)| \\
\label{eq:card_cup}
|\cross(X \cup Y)| &< |\cross(X)| \enspace .
\end{align}
By the definition of $X$, and as $\Fcal$ is closed under overlaps, we must have
$\chi(\Delta(X \cap Y)), \chi(\Delta(X \cup Y)) \in \laspan(\vec{\Lcal})$.
Also as $Y \in \Gcal$ we have $\chi(\Delta(Y)) \in \vec{\Lcal}$ which implies by \cref{prop:bixby} that
\[
\chi(\Delta(X)) =  \chi(\Delta(X \cap Y))+ \chi(\Delta(X \cup Y)) - \chi(\Delta(Y)) \enspace.
\]
This implies $\chi(\Delta(X)) \in \laspan(\vec{\Lcal})$, a contradiction.
\end{proof}

\end{document}